\documentclass[aps,prd,twocolumn,superscriptaddress]{revtex4-1}

\usepackage[pdftex]{graphicx}

\begin{document}


\title{Search for Galactic PeV Gamma Rays with the IceCube Neutrino Observatory}

\affiliation{III. Physikalisches Institut, RWTH Aachen University, D-52056 Aachen, Germany}
\affiliation{School of Chemistry \& Physics, University of Adelaide, Adelaide SA, 5005 Australia}
\affiliation{Dept.~of Physics and Astronomy, University of Alaska Anchorage, 3211 Providence Dr., Anchorage, AK 99508, USA}
\affiliation{CTSPS, Clark-Atlanta University, Atlanta, GA 30314, USA}
\affiliation{School of Physics and Center for Relativistic Astrophysics, Georgia Institute of Technology, Atlanta, GA 30332, USA}
\affiliation{Dept.~of Physics, Southern University, Baton Rouge, LA 70813, USA}
\affiliation{Dept.~of Physics, University of California, Berkeley, CA 94720, USA}
\affiliation{Lawrence Berkeley National Laboratory, Berkeley, CA 94720, USA}
\affiliation{Institut f\"ur Physik, Humboldt-Universit\"at zu Berlin, D-12489 Berlin, Germany}
\affiliation{Fakult\"at f\"ur Physik \& Astronomie, Ruhr-Universit\"at Bochum, D-44780 Bochum, Germany}
\affiliation{Physikalisches Institut, Universit\"at Bonn, Nussallee 12, D-53115 Bonn, Germany}
\affiliation{Universit\'e Libre de Bruxelles, Science Faculty CP230, B-1050 Brussels, Belgium}
\affiliation{Vrije Universiteit Brussel, Dienst ELEM, B-1050 Brussels, Belgium}
\affiliation{Dept.~of Physics, Chiba University, Chiba 263-8522, Japan}
\affiliation{Dept.~of Physics and Astronomy, University of Canterbury, Private Bag 4800, Christchurch, New Zealand}
\affiliation{Dept.~of Physics, University of Maryland, College Park, MD 20742, USA}
\affiliation{Dept.~of Physics and Center for Cosmology and Astro-Particle Physics, Ohio State University, Columbus, OH 43210, USA}
\affiliation{Dept.~of Astronomy, Ohio State University, Columbus, OH 43210, USA}
\affiliation{Dept.~of Physics, TU Dortmund University, D-44221 Dortmund, Germany}
\affiliation{Dept.~of Physics, University of Alberta, Edmonton, Alberta, Canada T6G 2G7}
\affiliation{D\'epartement de physique nucl\'eaire et corpusculaire, Universit\'e de Gen\`eve, CH-1211 Gen\`eve, Switzerland}
\affiliation{Dept.~of Physics and Astronomy, University of Gent, B-9000 Gent, Belgium}
\affiliation{Dept.~of Physics and Astronomy, University of California, Irvine, CA 92697, USA}
\affiliation{Laboratory for High Energy Physics, \'Ecole Polytechnique F\'ed\'erale, CH-1015 Lausanne, Switzerland}
\affiliation{Dept.~of Physics and Astronomy, University of Kansas, Lawrence, KS 66045, USA}
\affiliation{Dept.~of Astronomy, University of Wisconsin, Madison, WI 53706, USA}
\affiliation{Dept.~of Physics and Wisconsin IceCube Particle Astrophysics Center, University of Wisconsin, Madison, WI 53706, USA}
\affiliation{Institute of Physics, University of Mainz, Staudinger Weg 7, D-55099 Mainz, Germany}
\affiliation{Universit\'e de Mons, 7000 Mons, Belgium}
\affiliation{T.U. Munich, D-85748 Garching, Germany}
\affiliation{Bartol Research Institute and Department of Physics and Astronomy, University of Delaware, Newark, DE 19716, USA}
\affiliation{Dept.~of Physics, University of Oxford, 1 Keble Road, Oxford OX1 3NP, UK}
\affiliation{Dept.~of Physics, University of Wisconsin, River Falls, WI 54022, USA}
\affiliation{Oskar Klein Centre and Dept.~of Physics, Stockholm University, SE-10691 Stockholm, Sweden}
\affiliation{Department of Physics and Astronomy, Stony Brook University, Stony Brook, NY 11794-3800, USA}
\affiliation{Dept.~of Physics and Astronomy, University of Alabama, Tuscaloosa, AL 35487, USA}
\affiliation{Dept.~of Astronomy and Astrophysics, Pennsylvania State University, University Park, PA 16802, USA}
\affiliation{Dept.~of Physics, Pennsylvania State University, University Park, PA 16802, USA}
\affiliation{Dept.~of Physics and Astronomy, Uppsala University, Box 516, S-75120 Uppsala, Sweden}
\affiliation{Dept.~of Physics, University of Wuppertal, D-42119 Wuppertal, Germany}
\affiliation{DESY, D-15735 Zeuthen, Germany}

\author{M.~G.~Aartsen}
\affiliation{School of Chemistry \& Physics, University of Adelaide, Adelaide SA, 5005 Australia}
\author{R.~Abbasi}
\affiliation{Dept.~of Physics and Wisconsin IceCube Particle Astrophysics Center, University of Wisconsin, Madison, WI 53706, USA}
\author{Y.~Abdou}
\affiliation{Dept.~of Physics and Astronomy, University of Gent, B-9000 Gent, Belgium}
\author{M.~Ackermann}
\affiliation{DESY, D-15735 Zeuthen, Germany}
\author{J.~Adams}
\affiliation{Dept.~of Physics and Astronomy, University of Canterbury, Private Bag 4800, Christchurch, New Zealand}
\author{J.~A.~Aguilar}
\affiliation{D\'epartement de physique nucl\'eaire et corpusculaire, Universit\'e de Gen\`eve, CH-1211 Gen\`eve, Switzerland}
\author{M.~Ahlers}
\affiliation{Dept.~of Physics and Wisconsin IceCube Particle Astrophysics Center, University of Wisconsin, Madison, WI 53706, USA}
\author{D.~Altmann}
\affiliation{Institut f\"ur Physik, Humboldt-Universit\"at zu Berlin, D-12489 Berlin, Germany}
\author{K.~Andeen}
\affiliation{Dept.~of Physics and Wisconsin IceCube Particle Astrophysics Center, University of Wisconsin, Madison, WI 53706, USA}
\author{J.~Auffenberg}
\affiliation{Dept.~of Physics and Wisconsin IceCube Particle Astrophysics Center, University of Wisconsin, Madison, WI 53706, USA}
% ...


\date{\today}

\collaboration{IceCube Collaboration}
\noaffiliation




\begin{abstract}

Gamma-ray induced air showers are notable for their lack of muons, compared to hadronic showers. Hence, air shower arrays with large underground muon detectors can select a sample greatly enriched in photon showers by rejecting showers containing muons.   
IceCube is sensitive to muons with energies above $\sim$500 GeV at the surface, which provides an efficient veto system for hadronic air showers with energies above 1 PeV.
One year of data from the 40-string IceCube configuration was used to perform a search for point sources and a Galactic diffuse signal. No sources were found, resulting in a 90\% C.L. upper limit on the ratio of gamma rays to cosmic rays of $1.2\times10^{-3}$ 
for the flux coming from the Galactic Plane region ( $-80^\circ \lesssim l \lesssim -30^\circ$; $-10^\circ \lesssim b \lesssim 5^\circ$)
in the energy range 1.2 -- 6.0 PeV. In the same energy range, point source fluxes with $E^{-2}$ spectra have been excluded at a level of $(E/\mathrm{TeV})^2 \mathrm{d}\Phi/\mathrm{d}E \sim 10^{-12} - 10^{-11}$ cm$^{-2}$s$^{-1}$TeV$^{-1}$ depending on source declination. The complete IceCube detector will have a better sensitivity, due to the larger detector size, improved reconstruction and vetoing techniques. 
Preliminary data from the nearly-final IceCube detector configuration has been used to estimate the 5 year sensitivity of the full detector. It is found to be more than an order of magnitude better, allowing the search for PeV extensions of known TeV gamma-ray emitters.
 
\end{abstract}


\maketitle

\section{Introduction}

Gamma-rays are an important tool for studying the cosmos; unlike cosmic rays (CRs), they point back to their sources and can identify remote acceleration regions. Air Cherenkov telescopes have identified numerous sources of high-energy ($E > 1$ TeV) gamma-rays (see e.g. \cite{Aharonian:2008}): within our galaxy, gamma-rays have been observed coming from supernova remnants (SNRs), pulsar wind nebulae (PWNe), binary systems, and the Galactic Center.  Extra-galactic sources include starburst galaxies and Active Galactic Nuclei (AGNs).  Surface air-shower arrays like Milagro have performed all-sky searches for TeV gamma-rays.  Although these detectors are less sensitive to point sources than Air Cherenkov telescopes, they have identified several Galactic pointlike and extended sources  \cite{Abdo:2009}. 
Interactions of CRs with interstellar matter and radiation in the Galaxy produce a diffuse flux. Hadrons interacting with matter produce neutral pions, which decay into gamma rays, while CR electrons produce gamma-rays via inverse Compton scattering on the radiation field. Milagro has measured this diffuse Galactic flux in the TeV energy range with a median energy of 15 TeV and reported an excess in the Cygnus region, which might originate from CRs from local sources interacting with interstellar dust clouds \cite{Abdo:2008}.  
IceCube's predecessor AMANDA-II has also looked for TeV photons from a giant flare from SGR 1806-20,  using 100 GeV muons. AMANDA's large muon collection area compensated for the small cross-section for photons to produce muons \cite{Achterberg:2006az}.   

At higher energies, extra-galactic sources are unlikely to be visible, because more energetic photons are predicted to interact with cosmic microwave background radiation (CMBR), and with infrared starlight from early galaxies, producing $e^+e^-$ pairs \cite{Gould66}.  At 1 PeV, for example, photon propagation is limited to a range of about 10 kpc. It is unknown whether Galactic accelerators exist that can produce gamma rays of such high energy, but an expected flux results from interaction of (extragalactic) CRs with the interstellar medium (ISM) and dense molecular clouds.

To date, the best statistics on photons with energies in the range from $\sim$300 TeV to several PeVs come from the Chicago Air Shower Array - Michigan Muon Array (CASA-MIA), built at the Dugway Proving Ground in Utah.  CASA consisted of 1089 scintillation detectors placed on a square array with 15 m spacing.
MIA consisted of 1024 scintillation counters buried under about 3 m of earth, covering an area of 2500 m$^2$.  It served as a muon veto, with a threshold of about 0.8 GeV.    

CASA-MIA set a limit on the fraction of photons in the cosmic-ray flux of $10^{-4}$ at energies above 600 TeV \cite{Chantell:1997gs}.  The experiment also sets a limit of $2.4\times10^{-5}$ on the fraction of photons in the CR flux coming from within 5$^\circ$ of the galactic disk \cite{Borione:1997fy} at 310 TeV. This is near the theoretical expectation due to cosmic-ray interactions with the interstellar medium.  For a Northern hemisphere site like CASA-MIA, Ref.~\cite{Aharonian1991} predicts a gamma-ray fraction of $2\times10^{-5}$ for the average gas column density.

In this work, we present a new approach for detecting astrophysical PeV gamma rays, based on data of the surface component, IceTop, and the in-ice array of IceCube.
IceTop measures the electromagnetic component of air showers, while the in-ice array is sensitive to muons that penetrate the ice with energies above 500 GeV. 
While most CR showers above 1 PeV contain many muons above this threshold, only a small fraction of PeV gamma-ray showers carry muons that are energetic enough to reach the in-ice array.  
Therefore, gamma-ray candidates are selected among
muon-poor air showers detected with IceTop
and whose axis is reconstructed as passing through the in-ice array.

This approach of selecting muon-poor showers as gamma-ray candidates is fundamentally different from the earlier AMANDA-II gamma-ray search described above, which was only sensitive to gamma-ray showers that \emph{do} contain high energy muons ($> 100$ GeV).

We present a limit on the gamma-ray flux coming from the Galactic Plane, based on one year of data with half of the IceCube strings and surface stations installed.  We also discuss the sensitivity of the completed detector.

\section{The IceCube detector}
\label{sec:detector}

 \begin{figure}
 \includegraphics[width=\linewidth]{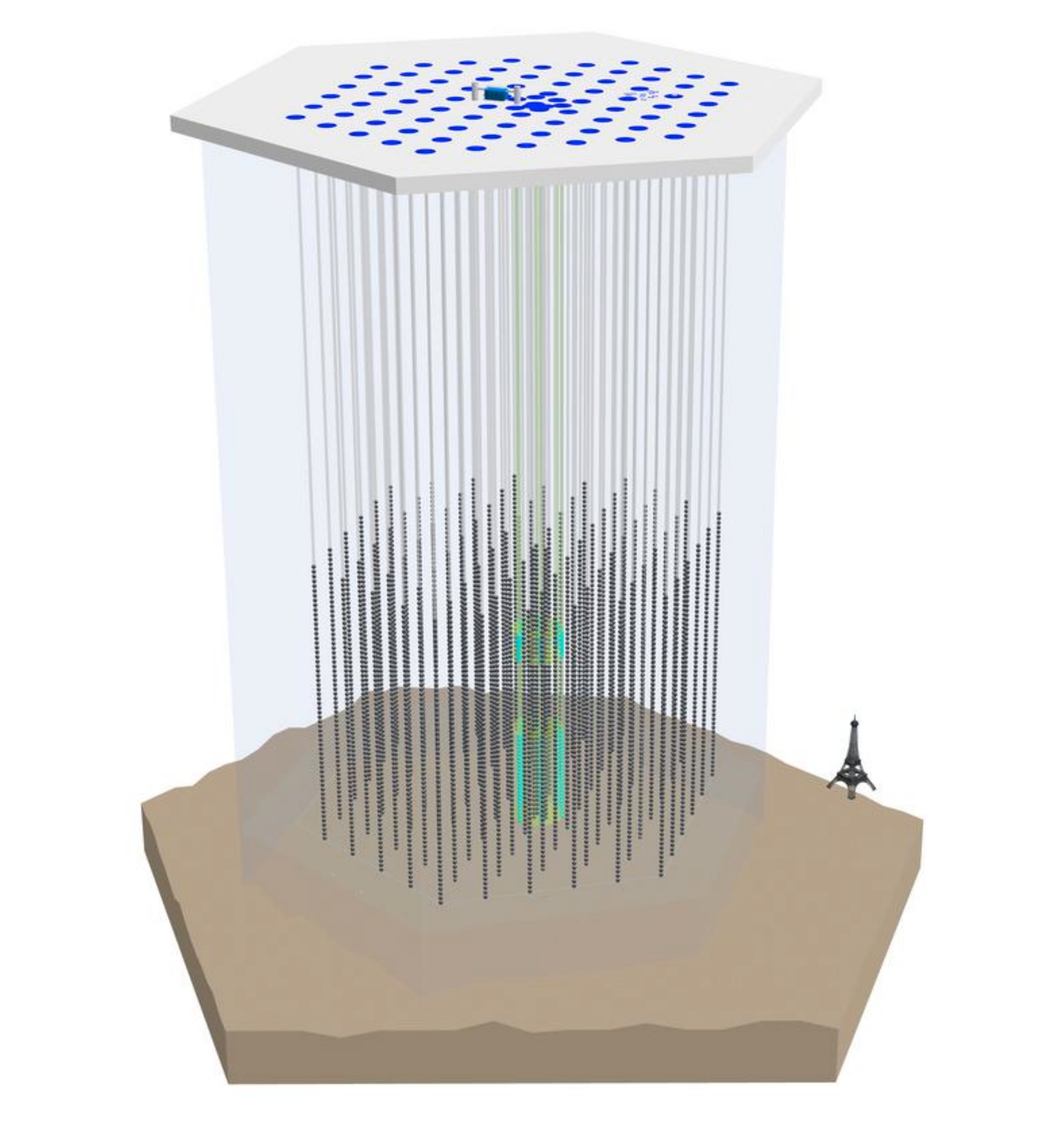}
 \caption{\label{fig:detector} IceCube consists of a $\sim$km$^2$ surface air shower array and 86 strings holding 60 optical modules each, filling a physical volume of a km$^{3}$. The region in the center of the buried detector is more densely instrumented. See text for details.}
 \end{figure}
 
\begin{figure}
 \includegraphics[width=\linewidth]{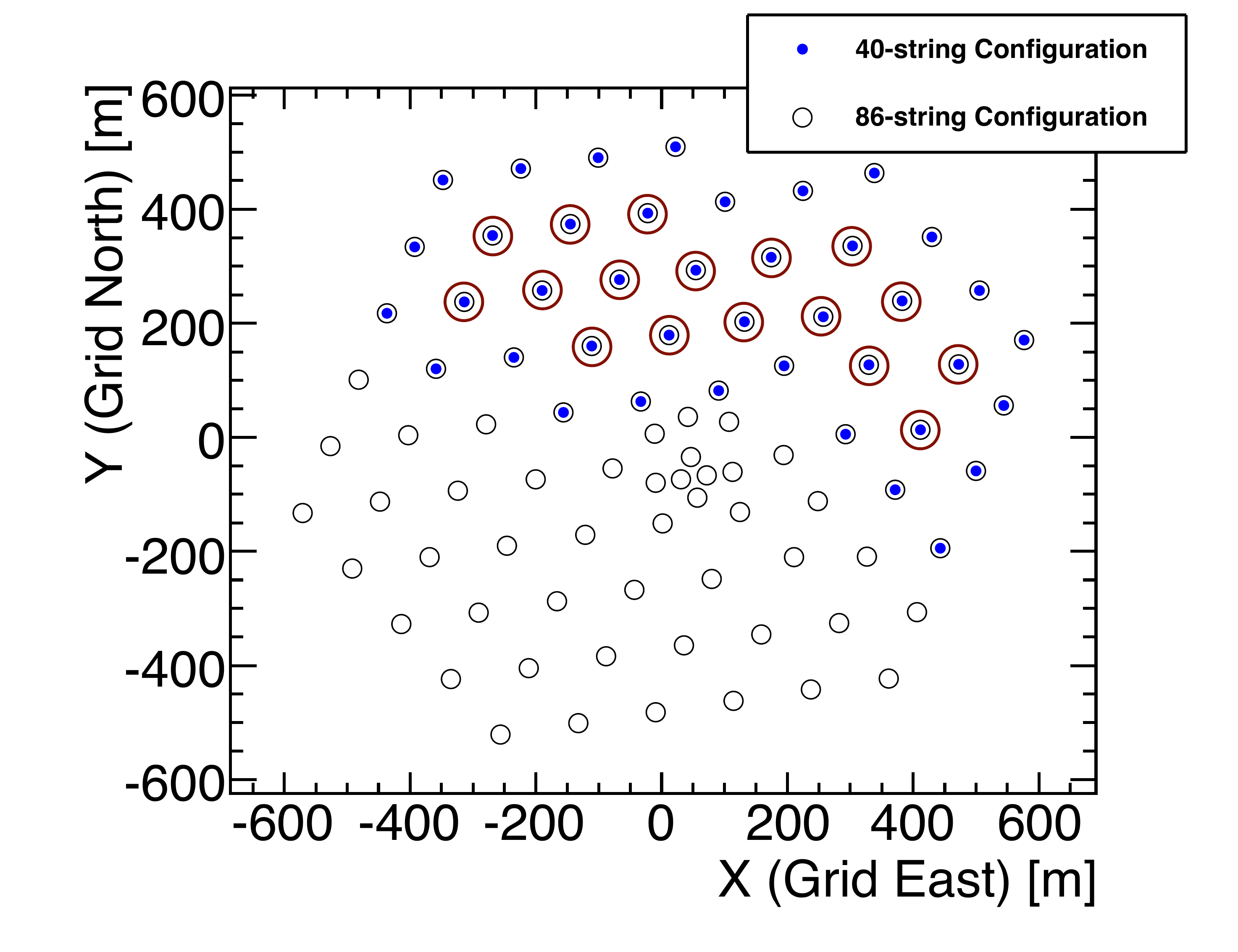}
 \caption{\label{fig:layout} Map of location of all 86 strings of the completed IceCube detector. The blue dots represent the 40 string configuration that is used for this analysis. At surface level each of these 40 strings is complemented by an IceTop station consisting of two tanks. The large (red) circles indicate the `inner strings' of the IC40 configuration.}
 \end{figure}

IceCube (see Fig.~\ref{fig:detector}) is a particle detector located at the geographic South Pole. The in-ice portion consists of 86 strings that reach 2450 m deep into the ice. Most of the strings are arranged in a hexagonal grid, separated by $\sim$125 m. Each of these strings holds 60 digital optical modules (DOMs) separated by $\sim$ 17 m covering the range from 1,450 m to 2,450 m depth. Eight strings form a denser instrumented area called DeepCore. The DOMs detect Cherenkov light produced by downward-going muons in cosmic-ray air showers and from charged particles produced in neutrino interactions. The data used in this analysis was collected in 2008/9, when the 40 strings shown in Fig.~\ref{fig:layout} were operational.

Each DOM is a complete detector system, comprising a 25 cm diameter Hamamatsu R7081-02 phototube 
\cite{Abbasi:2010vc}, shaping and digitizing electronics \cite{:2008ym}, calibration hardware, plus control electronics and power supply.   Most of the buried PMTs are run at a gain of $10^7$.  Digitization is initiated by a discriminator, with a threshold set to 0.25 times the typical peak amplitude of a single photoelectron waveform. Each DOM contains two separate digitizing systems; the Analog Transient Waveform Digitizer (ATWD) records 400 ns of data at 300 Megasamples/s, with a 14 bit dynamic range (divided among 3 parallel channels), while the fast Analog-to-Digital Converter (fADC) records 6.4 $\mu$s of data at 40 Megasamples/s, with 10 bits of dynamic range.  A system transmits timing signals between the surface and each DOM, providing timing calibrations across the entire array of about 2 ns \cite{Achterberg:2006md, Halzen:2010yj}.  

The IceTop surface array \cite{icetopdp} is located on the surface directly above the in-ice detectors.  It consists of 81 stations, each consisting of two ice-filled tanks, about 5~m apart.  For the 2008 data used here, 40 stations were operational (IC40, see Fig.\ \ref{fig:layout}).  
Each tank is 1.8 m in diameter, filled with ice to a depth of about 90 cm.  The tanks are initially filled with water, and the freezing of the water is controlled to minimize air bubbles and preserve the optical clarity of the ice.   Each tank is instrumented with two DOMs, a high-gain DOM run at a PMT gain of $5\times10^6$, and a low-gain DOM, with a PMT gain of $5\times 10^5$.  The two different gains were chosen to maximize dynamic range; the system is quite linear over a range from 1 to $10^5$ photoelectrons.  A station is considered hit when a low-gain DOM in one tank fires in coincidence with a high-gain DOM in the other; the thresholds are set to about 20 photoelectrons.

When an in-ice DOM is triggered it sends a Local Coincidence (LC) message to its nearest two neighbors above and nearest two neighbors below. If the DOM also receives an LC message from one of its neighbors within 1 $\mu$s it is in Hard Local Coincidence (HLC). In that case the full waveform information of both the ATWD and fADC chip is stored. For IC40 and earlier configurations, isolated or Soft Local Coincidence (SLC) hits were discarded. In newer configurations, the SLC hits are stored albeit with limited information. Keeping the full waveforms would require too much bandwidth, since the rate of isolated hits per DOM due to noise is $\sim$500 Hz. Instead, only the three fADC bins with the highest values and their hit times are stored.

In Sec.~\ref{sec:ic40ana} we present a gamma-ray analysis of the IC40 dataset. In Sec.~\ref{sec:fullsensitivity} we study the expected sensitivity of the full IceCube detector, and discuss how the inclusion of SLC hits increases the background rejection.

\section{Detection principle}
\label{sec:principle}

\begin{figure}
 \includegraphics[width=\linewidth]{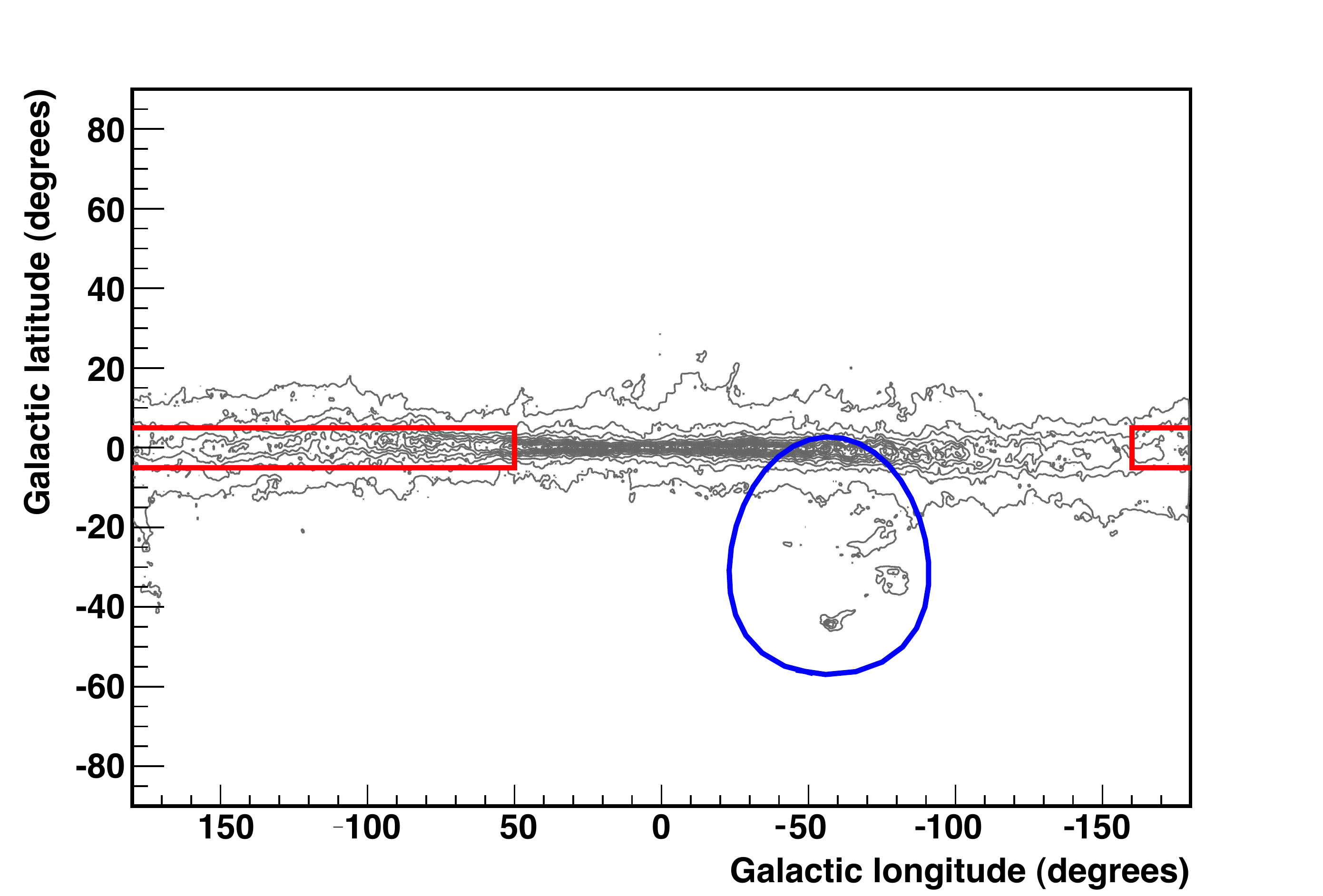}
 \caption{\label{fig:FOV} Contours of integrated neutral atomic hydrogen (HI) column densities \cite{LABsurvey}, in Galactic coordinates (flat projection). The blue circle indicates the gamma-ray FOV for IceCube in the present IC40 analysis. The red rectangle indicate the regions for which CASA-MIA \cite{Chantell:1997gs} has set an upper limit on the Galactic diffuse photon flux in the 100 TeV $-$ 1 PeV energy range. IceCube's FOV is smaller but covers a different part of the Galactic Plane, close to the Galactic Center.}
 \end{figure}

We create a sample of gamma-ray candidates by selecting air showers that have been successfully reconstructed by IceTop and have a shower axis that passes through the IceCube instrumented volume. The geometry limits this sample to showers having a maximum zenith angle of $\sim$30 degrees. Since IceCube is located at the geographic South Pole, the Field of View (FOV) is roughly constrained to the declination range of $-60$ to $-90$ degrees, as shown in Fig.~\ref{fig:FOV}.  This FOV includes the Magellanic Clouds and part of the Galactic Plane. Gamma-rays at PeV energies are strongly attenuated over extra-galactic distances, thus limiting the observable sources to those localized in our galaxy. At distances of $\sim$ 50~kpc and $\sim$60~kpc, the PeV gamma-ray flux from the Large and Small Magellanic Cloud is suppressed by several orders of magnitude.

The contours in the background of Fig.~\ref{fig:FOV} are the integrated neutral atomic hydrogen (HI) column densities under the assumption of optical transparency based on data from the Leiden/Argentine/Bonn survey \cite{LABsurvey}. These densities are not incorporated into the analysis and are only plotted to indicate the Galactic Plane. We do expect, however, that gamma-ray sources are correlated with the HI column density. Firstly, Galactic CR accelerators are more abundant in the high density regions of the Galaxy. Secondly, the gamma-ray flux of (extra-)galactic CRs interacting with the ISM naturally correlates with the column density. However, it has to be noticed that this correlation is not linear, because of the attenuation of gamma-rays over a 10 kpc distance scale.
Furthermore, the column densities do not include molecular hydrogen which can also act as a target for CRs.

The gamma-ray candidate events are searched for in a background of CR showers that have exceptionally few muons or are directionally misreconstructed. In the latter case the muon bundle reaches kilometers deep into the ice but misses the instrumented volume. This background is hard to predict with Monte Carlo (MC) simulations. Cosmic-ray showers at PeV energies and with a low number of energetic muons are rare. For example, at 1 PeV less than 0.1\% of the simulated showers contain no muons with an energy above 500 GeV, approximately what is needed to reach the detector in the deep ice when traveling vertically.  Determining how many hadronic showers produce a signal in a buried DOM would require an enormous amount of MC data to reach sufficient statistics, plus very strong control of the systematic uncertainties due to muon production, propagation of muons and Cherenkov photons through the ice, and the absolute detector efficiencies.
It would also have to be able to accurately predict the errors in air shower reconstruction parameters. For example, this analysis is very sensitive to the tails of the distribution of the error on the angular reconstruction of IceTop. Even MC sets that are large enough to populate these tails are not expected to properly describe them.

To avoid these issues, we determine the background directly from data. As a result, we are not able to measure a possible isotropic contribution to the gamma-ray flux, because these gamma-rays would be regarded as background. Instead, we search for localized excesses in the gamma-ray flux.   

We search for a correlation of the arrival directions of the candidate events with the Galactic Plane, and scan for point sources. 
The acceptance of IceTop-40 is a complex function of azimuth and zenith due to its elongated shape and the requirement that the axis of the detected shower passes through IC40 (with the same elongated shape). 
However, since the arrival time is random (there are no systematic gaps in detector uptime w.r.t.\ sidereal time) 
the reconstructed right ascension (RA) of an isotropic flux of CR showers is uniform.
The correct declination distribution of the background is very sensitive to the number of background showers introduced by errors in the IceTop angular reconstruction of the air shower as a function of the zenith angle, and is taken to be unknown. However, the flat distribution of background over RA is enough to allow for a search for gamma-ray sources.

Recently, IceCube found an anisotropy in the arrival direction of CRs on the southern hemisphere \cite{ic:anisotropy}. 
These deviations with RA have been established for samples of CRs with median energies of 20 TeV and 400 TeV. 
The two energy ranges show a very different shape of the anisotropy, but the level of the fractional variations in flux is at a part-per-mil level for both \cite{ic:anisotropy2}.
An anisotropy with comparable magnitude in the PeV energy range is too small to affect this analysis (the IC40 final sample contains 268 events). 

\section{IC40 analysis}
\label{sec:ic40ana}
\subsection{Event selection}
Between April 5 2008 and May 20 2009, IceCube took data with a configuration of 40 strings and 40 surface stations (IC40), using several trigger conditions based on different signal topologies. This analysis uses the 8 station surface trigger, which requires a signal above threshold in both tanks of at least 8 IceTop stations. An additional signal in IceCube is not required for this trigger, but all HLC hits in the deep detector within a time window of 10 $\mu s$ before and after the surface trigger are recorded. 

The air shower parameters are reconstructed from the IceTop hits with a series of likelihood maximization methods.
The core position is found by fitting the lateral distribution of the signal, using
\begin{equation}
S(r) = S_{\mathrm{ref}} \left(\frac{r}{R_{\mathrm{ref}}}\right)^{-\beta-\kappa \log_{10}\left(\frac{r}{R_{\mathrm{ref}}}\right)}
\label{eq:latfit}
\end{equation}
where $S$ is the signal strength, $r$ is the distance to the shower axis,  $S_{\mathrm{ref}}$ is the fitted signal strength at the reference distance $R_{\mathrm{ref}}=125$ m, $\beta$ is the slope parameter reflecting the shower age, and $\kappa=0.303$ is a constant  determined from simulation \cite{icetopdp,IceCube:2012wn}. Signal times are used to find the arrival direction of the air shower. The time delay due to the shape of the shower plane is described by the sum of a Gaussian function and a parabola, both centered at the shower core, which yields a resolution of $1.5^{\circ}$ for IC40. 
The relationship between the reconstructed energy $E_{\mathrm{reco}}$ and $S_{\mathrm{ref}}$ is based on MC simulations for proton showers and depends on the zenith angle. 
 
IceCube data is processed in different stages: in the first two levels the raw data is calibrated and filtered, and various fitting algorithms are applied, of which only the IceTop reconstruction described above is used in this analysis. In the selection of photon shower candidates from the data sample we distinguish two more steps: level three (L3) and level four (L4). Level three includes all the conditions on reconstruction quality, geometry and energy that make no distinction between gamma-ray showers and CR showers. The L4 cut is designed to separate gamma rays from CRs.

Two parameters are used to constrain the geometry and ensure the shower axis passes through the instrumented volume of IceCube. The IceTop containment parameter $C_{\mathrm{IT}}$ is a measure for how centralized the core location is in IceTop. When the core is exactly in the center of the array $C_{\mathrm{IT}}=0$, while $C_{\mathrm{IT}}=1$ means that it is exactly on the edge of the array. More precisely, $C_{\mathrm{IT}}=x$ means that the core would have been on the edge of the array if the array would be $x$ times its actual size. The string distance parameter $d_{\mathrm{str}}$ is the distance between the point where the shower reaches the depth of the first level of DOMs and the closest inner string. Inner string, in this sense, means a string which is not on the border of the detector configuration. IC40 has 17 inner strings (see Fig.~\ref{fig:layout}). The L3 cuts are:
\begin{itemize}
\item Quality cut on lateral distribution fit: $1.55 < \beta < 4.95$ (cf. Eq.~\ref{eq:latfit})
\item Geometry cut: $C_{\mathrm{IT}} < C_{\mathrm{max}}$
\item Geometry cut: $d_{\mathrm{str}} < d_{\mathrm{max}}$
\item Energy cut: $E_{\mathrm{reco}} > E_{\mathrm{min}} $
\end{itemize}
The energy and geometry cuts are optimized in a later stage (Sec.~\ref{sec:gptest}).

The L4 stage imposes only one extra criterion to the event: there should be no HLC hits in IceCube. This removes most of the CR showers, if $E_{\mathrm{min}}$ is chosen sufficiently high. The remaining background consists of CR showers with low muon content and mis-reconstructed showers, the high energy muons from which do not actually pass through IceCube. 

The event sample after the L4 cut might be dominated by remaining background but it can be used to set an upper limit to the number of gamma-rays in the sample. Since the event sample after L3 cuts is certainly dominated by CRs, the ratio between the number of events after L4 and L3 cuts can be used to calculate an upper limit to the ratio of gamma-ray-to-CR showers.

The remaining set of candidate gamma ray events is tested for a correlation with the Galactic Plane (Sec.~\ref{sec:gptest}) and the presence of point like sources (Sec.~\ref{sec:pssearch}). First, the results of simulations are presented, which provide several quantities needed for sensitivity calculations.      

\subsection{Simulation}
\label{sec:mc}
Although we determine the background from data only, simulations are needed to investigate systematic differences in the detector response to gamma-ray showers and cosmic-ray showers. More specifically, we are interested in the energy reconstruction of gamma-ray showers, the fraction of gamma-ray showers that is rejected by the muon veto system, and a possible difference in effective detector area between both types of showers.

Gamma-ray and proton showers are simulated with CORSIKA v6.900, using the interaction models FLUKA 2008 and SYBILL 2.1 for  low and high energy hadronic interactions, respectively.  For both primaries 10,000 showers are generated within an energy range of 800 TeV to 3 PeV with a $E^{-1}$ spectrum. Because the shower axes are required to pass through IceCube, the zenith angle is restricted to a maximum of 35 degrees. The observation altitude for IceTop is 2835~m. Atmospheric model MSIS-90-E is used, which is South Pole atmosphere for July 01, 1997. Seasonal variations in the event rate are of the order of 10\% \cite{tilav09}.

The detector simulation is done with the IceTray software package. Each simulated shower is fed into the detector simulation ten times with different core positions and azimuthal arrival direction, for a total of 100,000 events for both gamma rays and protons. This resampling of showers is a useful technique for increasing the statistics when examining quantities like the resolution of the energy reconstruction of IceTop.
However, it cannot be used for quantities with large shower-to-shower variations, such as the number of high energy muons.  

 \begin{figure}
 \includegraphics[width=\linewidth]{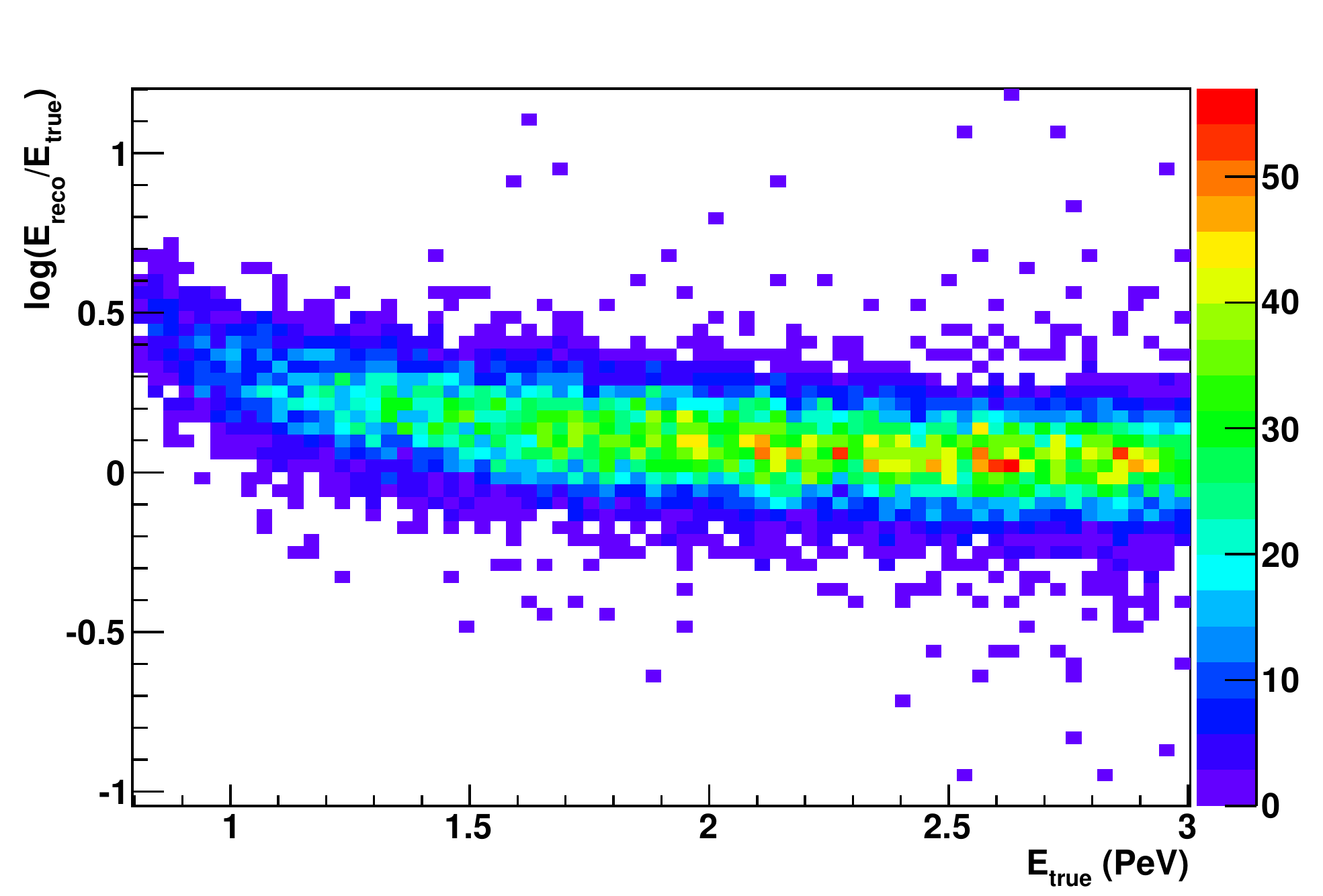}
 \caption{\label{fig:Ereco} Ratio between reconstructed and true energy of simulated gamma-ray showers as a function of their true energy. At low energies the overestimation of the gamma-ray energy is largely due to a bias effect of the eight-station filter. At higher energies, this overestimation decreases. }
 \end{figure}
 
 \begin{figure}
 \includegraphics[width=\linewidth]{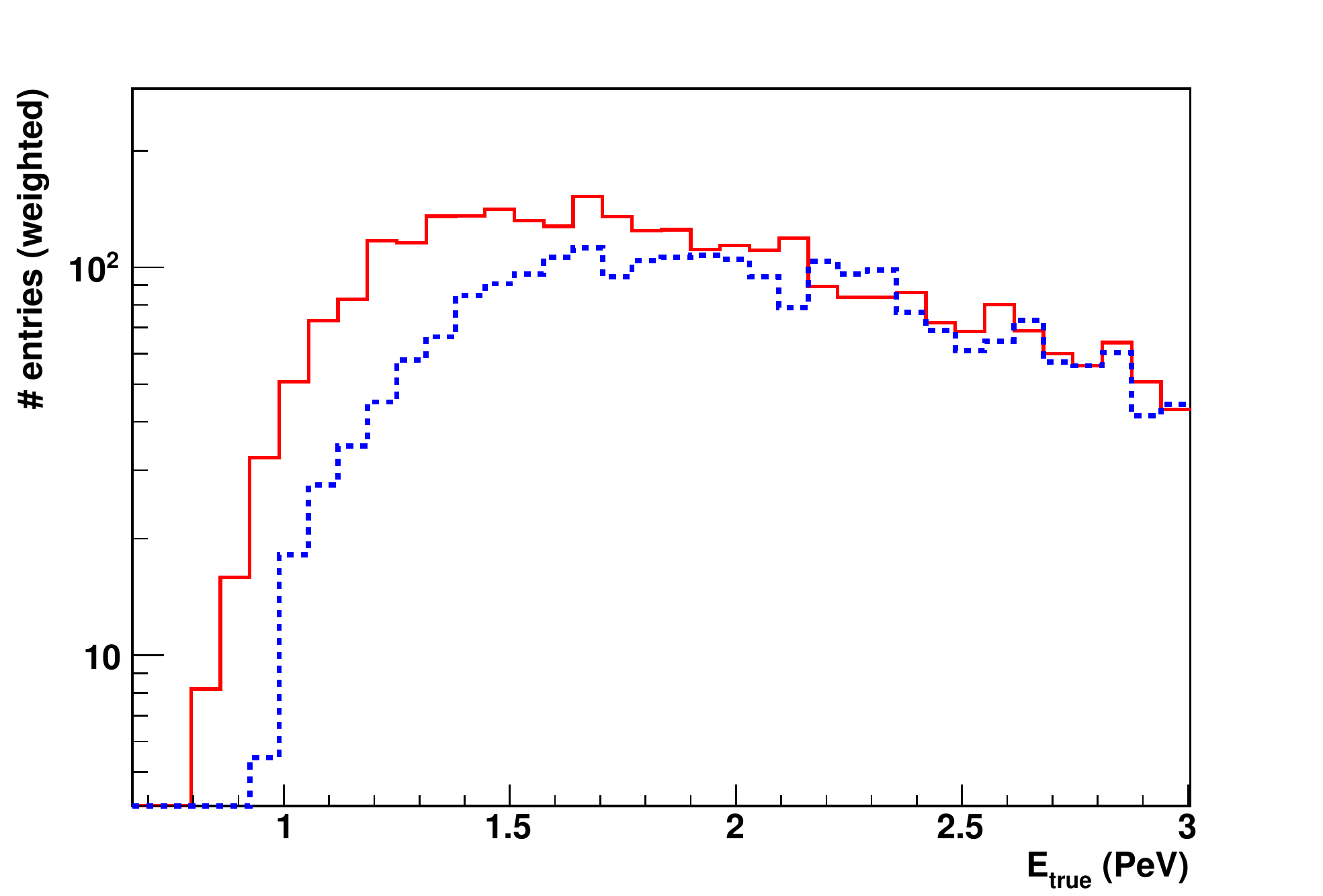}
 \caption{\label{fig:Ethresh} Distribution of true energy of gamma-ray (red, solid) and proton (blue, dotted) showers for an energy cut at $E_{\mathrm{reco}}>1.4$~PeV, weighted to a $E^{-2.7}$ spectrum.}
 \end{figure}
 
The energy of gamma-ray showers is overestimated by the reconstruction algorithm. Fig.~\ref{fig:Ereco} shows the distribution of the logarithm of the ratio between the reconstructed and true primary energy as function of true energy, weighted to a $E^{-2.7}$ spectrum. There are two reasons for the energy offset. First, there is a selection effect of the eight-station filter, which has a bias (below a few PeV) towards showers that produce relatively large signals in the IceTop tanks. This also affects the reconstructed energy of CR showers. At higher energies, the offset decreases but the reconstructed energy of gamma-ray showers is still slightly overestimated because the energy calibration of IceTop is performed with respect to proton showers.

Figure \ref{fig:Ethresh} shows the distribution of true energies of gamma-ray and proton showers for the energy cut $E_{\mathrm{reco}}>1.4$~PeV (which will be adopted in Sec.~\ref{sec:gptest}). After this cut, 95\% of the gamma-ray showers have a true energy above 1.2 PeV, while 95\% of the proton showers have an energy above 1.3 PeV.     

The fraction of gamma-ray showers that is falsely rejected because the showers contain muons that produce a signal in IceCube is found by applying the cuts to the MC simulation. 

After applying the L3 cuts (defined in Sec.~\ref{sec:gptest}) to the simulated gamma-ray sample there are 737 events left in the sample, of which 121 produce a signal in IceCube. Taking into account an energy weighting of $E^{-2.7}$, this corresponds to 16\%. 

Showers that have no energetic muons can still be rejected if an unrelated signal is detected by IceCube. This could be caused by noise hits or unrelated muon tracks that fall inside the time window. This noise rate is determined directly from the data and leads to 14\% signal loss. The total fraction of gamma-ray showers that is falsely rejected is therefore 28\%. 

Finally, because the composition, shower size and evolution of gamma-ray and CR showers are different, one might expect a difference in the number of triggered stations and the quality of the reconstruction, which could lead to different effective areas. Such an effect would be of importance when calculating the relative contribution of gamma-rays to the total received flux.  
We compare the effective area for gamma rays and protons by counting the number of events that are present at L3. We compensate for the energy reconstruction offset by reducing the reconstructed gamma-ray energy by a factor 1.16. The ratio of the effective area for gamma-rays to that for protons is then found to be 0.99.

It should be emphasized that we do not use the simulation to determine the number of muons and their energy distribution from CR showers. This would require a simulation set that includes heavier nuclei instead of protons only.  Moreover, various hadronic interaction models generate significantly different muon fluxes \cite{pierog06}. Instead, this analysis estimates the rate of CR showers that do not trigger IceCube using the data itself.   

\subsection{Galactic Plane test}
\label{sec:gptest}

The IC40 data set consists of 368 days of combined IceCube and IceTop measurements. The data from August 2009 is used as a burn sample, which means that it is used to tune the parameters of the analysis. After this tuning the burn sample is discarded and the remaining data is used for the analysis. 

IceCube is sensitive to gamma rays above 1 PeV.
Earlier searches by CASA-MIA in a slightly lower energy range (100 TeV -- 1 PeV) with better sensitivity have not established a correlation of gamma-rays with the Galactic Plane (see Fig.~\ref{fig:sensgalp}).
For a Galactic diffuse flux below the CASA-MIA limit \cite{Chantell:1997gs} no gamma rays are expected in the IC40 burn sample.  
However, IceCube observes a different part of the Galactic Plane (see Fig.~\ref{fig:FOV}), close to the Galactic Center, so the possibility exists that previously undetected sources or local enhancements in CR and dust densities create an increase in the flux from that part of the sky.

In order to find a possible correlation of candidate events in our burn sample to the Galactic Plane, different sets of L3 cut parameters are applied to find a set that produces the most significant correlation. Afterwards, the cut parameters are fixed and the burn sample is discarded. The fixed cut parameters are then used in the analysis of the rest of the IC40 data set to test whether the correlation is still present. Note that these cuts are applied at L3, so they affect both the event samples after L3 and L4 cuts. This is important because the ratio between the number of events after L4 and L3 cuts is used to calculate the ratio between gamma ray and CR showers.

There are three cut parameters that are optimized by using the burn sample: $E_{\mathrm{min}}$, $C_{\mathrm{max}}$ and $d_{\mathrm{max}}$. This is done by scanning through all combination of parameters within the following range:  600 TeV $ \leq E_{\mathrm{min}}  \leq  $ 2 PeV with steps of 100 TeV, $0.5 \leq C_{\mathrm{max}} \leq 1.0$ with steps of 0.1, and 50 m $\leq d_{\mathrm{max}} \leq 90$ m  with steps of 10 m. 
For each combination, the number of events $N_S$ in the burn sample after the L4 cut, located in the source region, is counted. 
The source region is defined as within 10 degrees of the Galactic Plane. Then, the data set is scrambled multiple times by randomizing the RA of each data point. For each scrambled data set the number of events in the source region is again counted. 
The best combination of cut parameters is the set which has the lowest fraction of scrambled data sets for which this number is equal to or higher than $N_S$. 

 \begin{figure*}
 \includegraphics[width=0.32\linewidth]{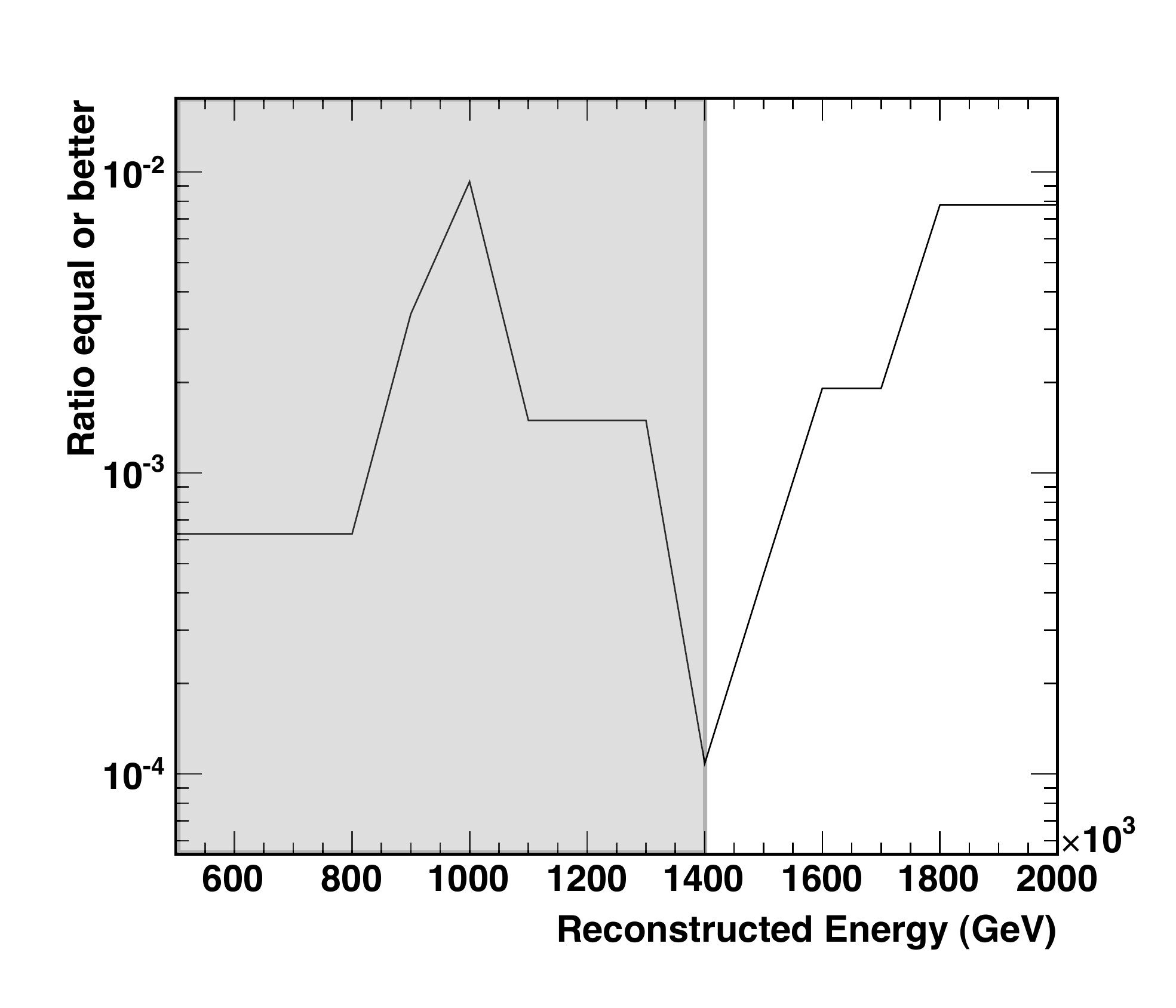}
 \includegraphics[width=0.32\linewidth]{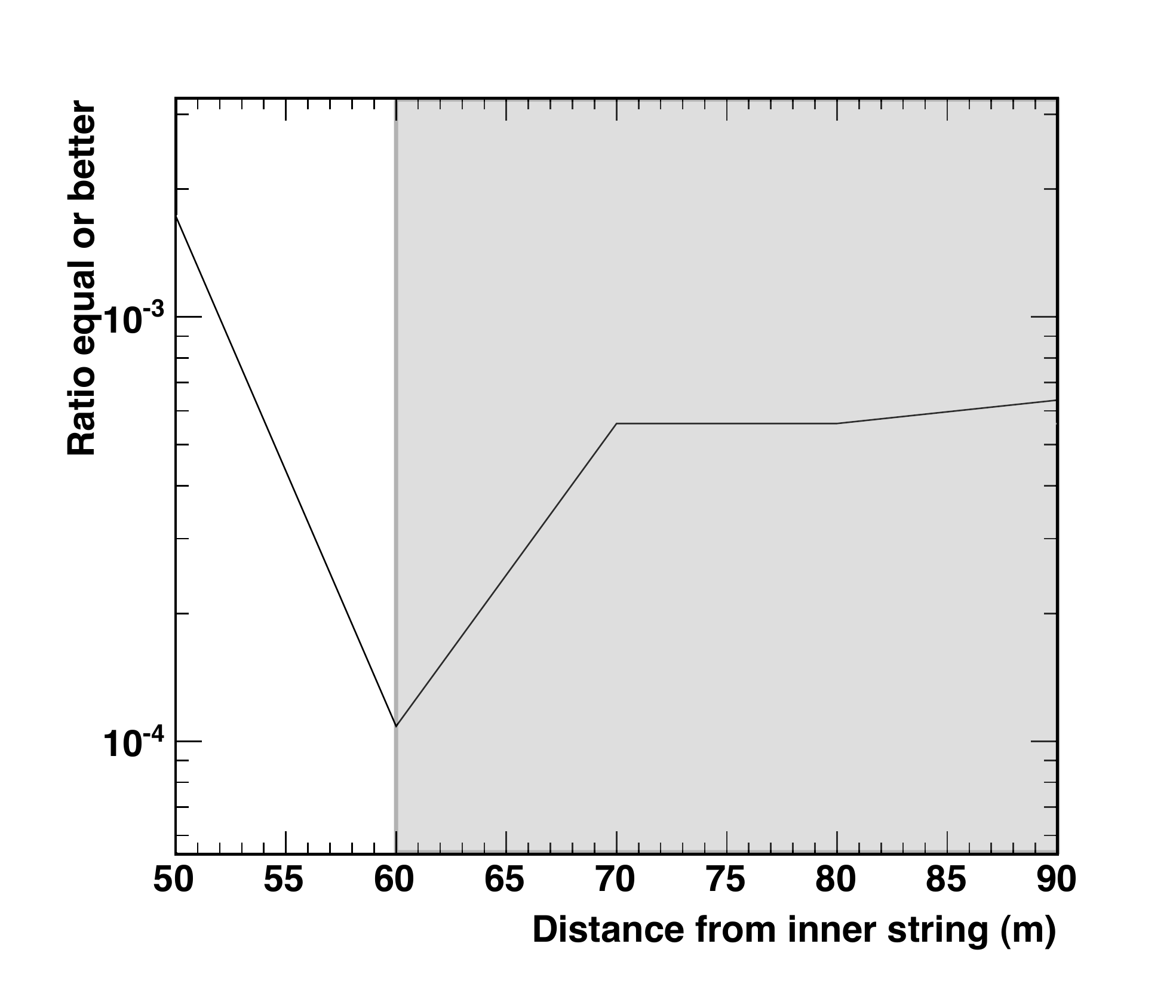}
 \includegraphics[width=0.32\linewidth]{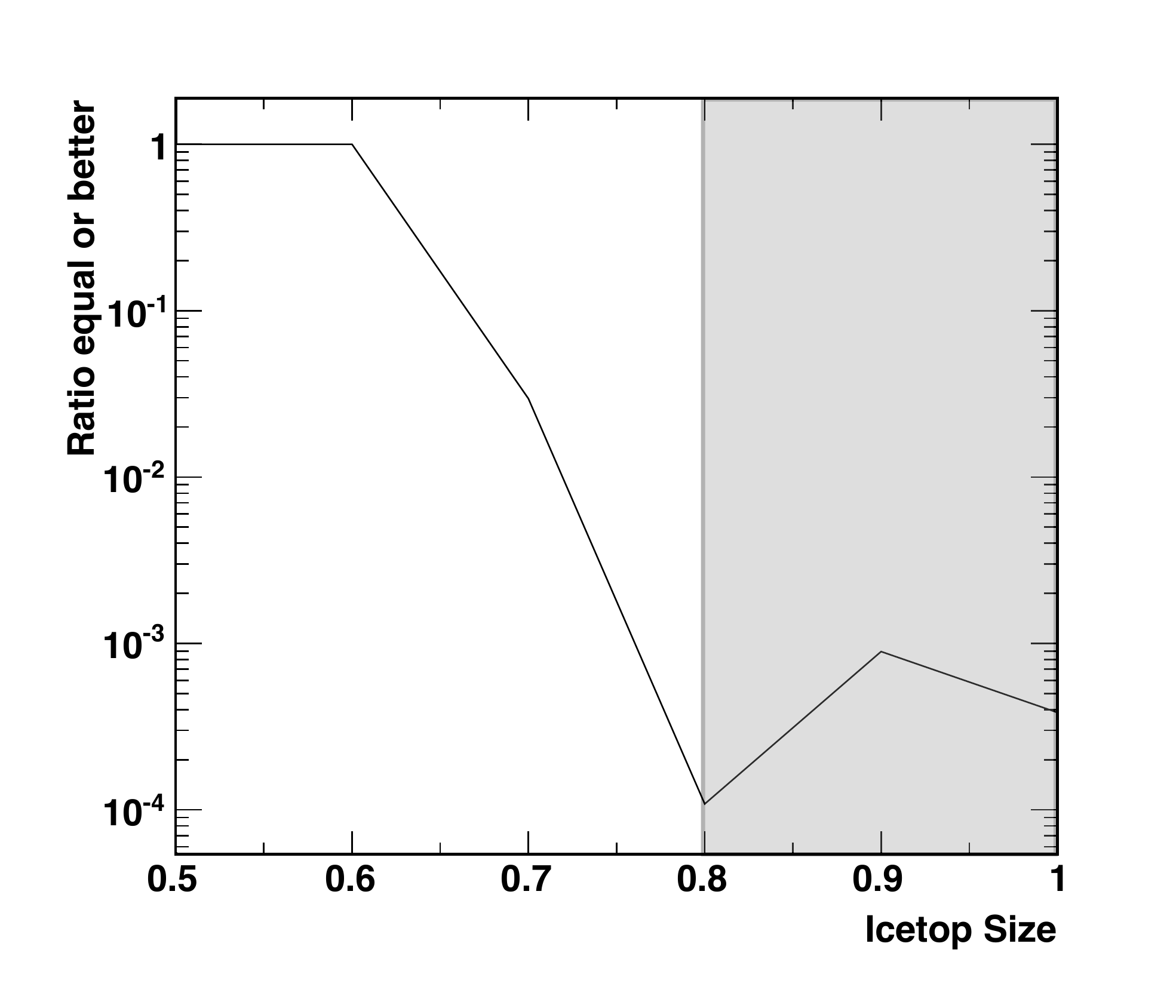}
 \caption{\label{fig:optim} Optimization scans for cut parameters $E_{\mathrm{min}}$, $d_{\mathrm{max}}$, and $C_{\mathrm{max}}$. The fraction of scrambled data sets that perform equal or better than the real data set is plotted against cut value. For each plot, the other 2 parameters are kept constant at their optimal value. The actual scan was done three dimensionally. For each plot, the shaded region indicates the parameter space that is excluded by the optimized cuts.}
 \end{figure*}

The result of the scan is given in the three panels of Fig.~\ref{fig:optim}. For each cut parameter the fraction of scrambled data sets that has a number of events in the source region equal to or exceeding the amount in the original data set is plotted for different cut values. For each plot the values of the other two cut parameters are kept constant at their optimal value. The actual search is done in three dimensions. The ratio is lowest for $E>1.4$ PeV, $C_{\mathrm{IT}}<0.8$ and $d_{\mathrm{str}}<60$ m. With this combination of cuts only 0.011\% of the scrambled data sets produce an equal or higher number of events in the source region. 

Note that while this procedure of optimizing cuts should be effective in the presence of sufficient signal, the small fraction obtained here and its erratic behavior with changing cut values are consistent with fluctuations of the background CR distribution alone, given the large number of possible cut combinations which were scanned. Nonetheless, the cut parameter values found with this procedure seem very reasonable (similar values are found with an alternative method, see Sec.~\ref{sec:fullsensitivity}).

 \begin{figure*}
 \includegraphics[width=\linewidth]{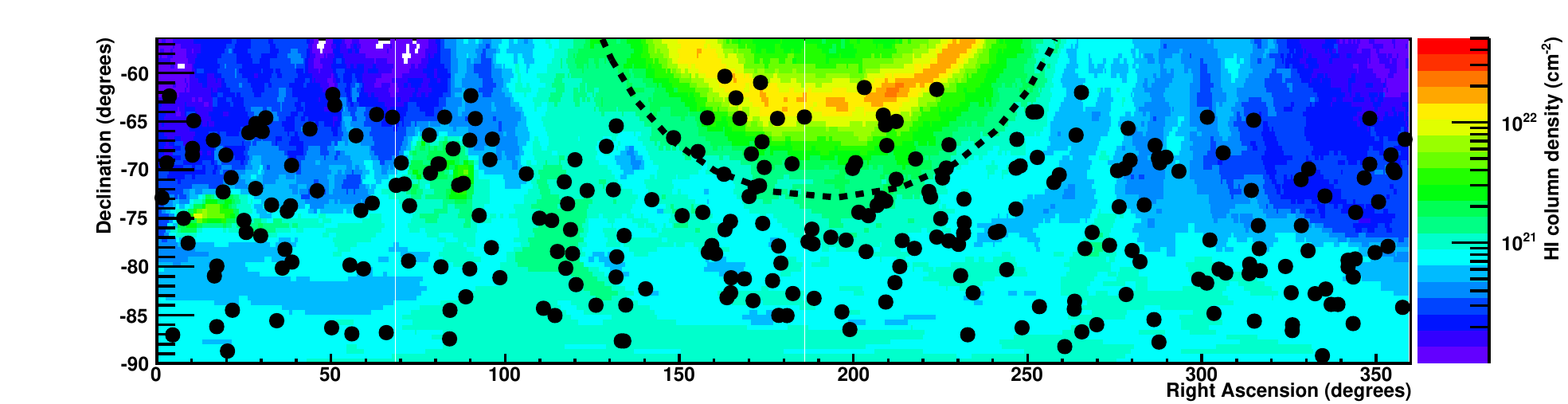}
 \caption{\label{fig:skymap} Equatorial map of the 268 candidate gamma-ray events of the IC40 data set superimposed on HI column densities based on \cite{LABsurvey}. The dotted black curve encloses the source region, defined as within 10 degrees of the Galactic Plane.}
 \end{figure*}
 
The optimized cuts are applied to the complete IC40 data set minus the burn sample. There are 268 candidate events of which 28 are located in the source region. Figure \ref{fig:skymap} is a map of the sky showing all 268 events. 
The colors in the background indicate the integrated HI column densities, cf. Fig.~\ref{fig:FOV} (see discussion Sec.~\ref{sec:principle}). These are meant to guide the eye and are not part of the analysis.

The significance of the correlation with the Galactic Plane is tested by producing data sets with scrambled RA. An equal or higher number of source region events is found in 21\% of the scrambled data sets, corresponding to a non-significant excess of +0.9$\sigma$.

We follow the procedure of Feldman \& Cousins \cite{FC98} to construct an upper limit for the ratio of gamma rays to CRs. The background is determined by selecting a range of RA that does not contain the source region. Within this range the data points are scrambled multiple times and for each scrambled set the number of events in a pre-defined region of the same shape and size as the source region is counted. This yields a mean background of 24.13 events for the source region. Using a 90\% confidence interval, the upper limit on the number of excess gamma rays from this region is 14. 
 
Since 28\% of gamma-ray showers are rejected by the veto from the buried detector, the maximum number of excess gamma-rays from the Galactic Plane is $14/0.72=19.4$. From Fig.~\ref{fig:Ethresh}, it is known that the energy cut corresponds to a threshold of 1.2 PeV for gamma-rays, and 1.3 PeV for protons. Given that at L3 the sample is dominated by CR showers, and assuming a CR and gamma-ray power law of $\gamma=-2.7$, a 90\% C.L. upper limit of $1.2\times 10^{-3}$ on the ratio of gamma-ray showers to CR showers in the source region can be derived in the energy range 1.2 -- 6.0 PeV.
The upper bound of 6.0 PeV is the value for which 90\% of the events are inside the energy range. This value falls outside the range for which gamma-ray showers were simulated. However, there is no indication that the energy relation plotted in Fig.~\ref{fig:Ereco} behaves erratically above 3 PeV.    

This is a limit on the average excess of the ratio of gamma-rays to CRs in the source region with respect to the rest of the sky, i.e.\ a limit on the Galactic component of the total gamma-ray flux. A possible isotropic component is not included. 
Systematic uncertainties lead to a 18\% variation of the upper limit, as determined in Sec.~\ref{sec:systerror}.

\subsection{Unbinned point source search}
\label{sec:pssearch}

An additional search for point-like sources tests the possibility that a single source dominates the PeV gamma-ray sky.   This source does not necessarily lie close to the Galactic Plane. An unbinned point source search is performed on the sky within the declination range of $-85^{\circ}$ to $-60^{\circ}$, using a method that follows \cite{Braun08}. The region within 5 degrees around the zenith is omitted, because the method relies on scrambled data sets that are produced by randomizing the RA of the events. Close to the zenith this randomization scheme fails due to the small number of events.  

The data is described by an unknown amount of signal events on top of a flat distribution of background events. In an unbinned search, a dense grid of points in the sky is scanned. For each point a maximum likelihood fit is performed for the relative contribution of source events over background events.

 \begin{figure*}
 \includegraphics[width=\linewidth]{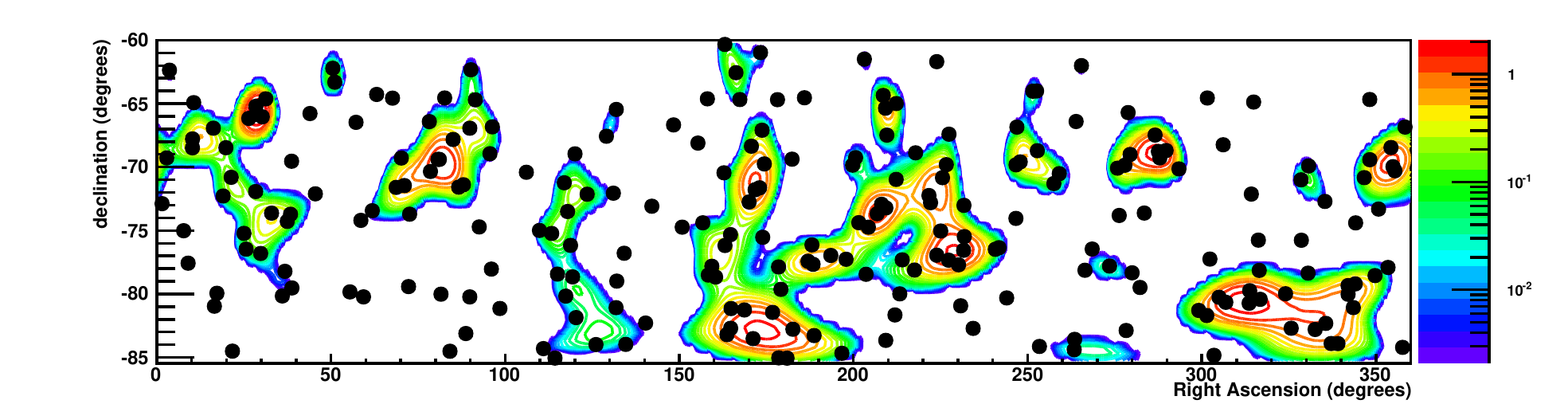}
 \caption{\label{fig:pointsources} Equatorial map of the part of the sky for which an unbinned point source search is performed. The contours indicate the value of $\lambda$ and the black dots are the candidate events. }
 \end{figure*}
 
For a particular event $i$ the probability density function (PDF) is given by
\begin{equation}
P_i(n_S)=\frac{n_S}{N} S_i + \left( 1- \frac{n_S}{N} \right) B_i
\end{equation}
where $n_S$ is the number of events that is associated to the source, $B_i$ is the background PDF, and
\begin{equation}
S_i= \frac{1}{2 \pi \sigma^2} \exp\left(-\frac{\Delta \Psi^2}{2 \sigma^2}\right)
\end{equation}
is the two-dimensional Gaussian source PDF, in which $\Delta\Psi$ is the space angle between the event and the source test location, and $\sigma=1.5^{\circ}$ is the angular resolution of IceTop. The background PDF $B_i$ is only dependent on the zenith angle, and is derived from the zenith distribution of the data. For each point in the sky there is a likelihood function
\begin{equation}
L(n_S)=\Pi_i P_i(n_S),
\end{equation}
and associated test statistic
\begin{equation}
\lambda= -2 \left( \log(L(0)) - \log(L(n_S) \right),
\end{equation}
which is maximized for $n_S$. In the optimization procedure, $n_S$ is allowed to have a negative value, which mathematically corresponds to a local flux deficit. 
 
The procedure is similar to the search method for the neutrino point sources with IceCube \cite{icecube:ps}, except that the source and background PDF do not contain an energy term. 
Because the range of energies in the event sample is relatively small (90\% of the events have an energy between 1 and 6 PeV), an energy PDF is unlikely to improve the sensitivity.

  \begin{figure}
 \includegraphics[width=\linewidth]{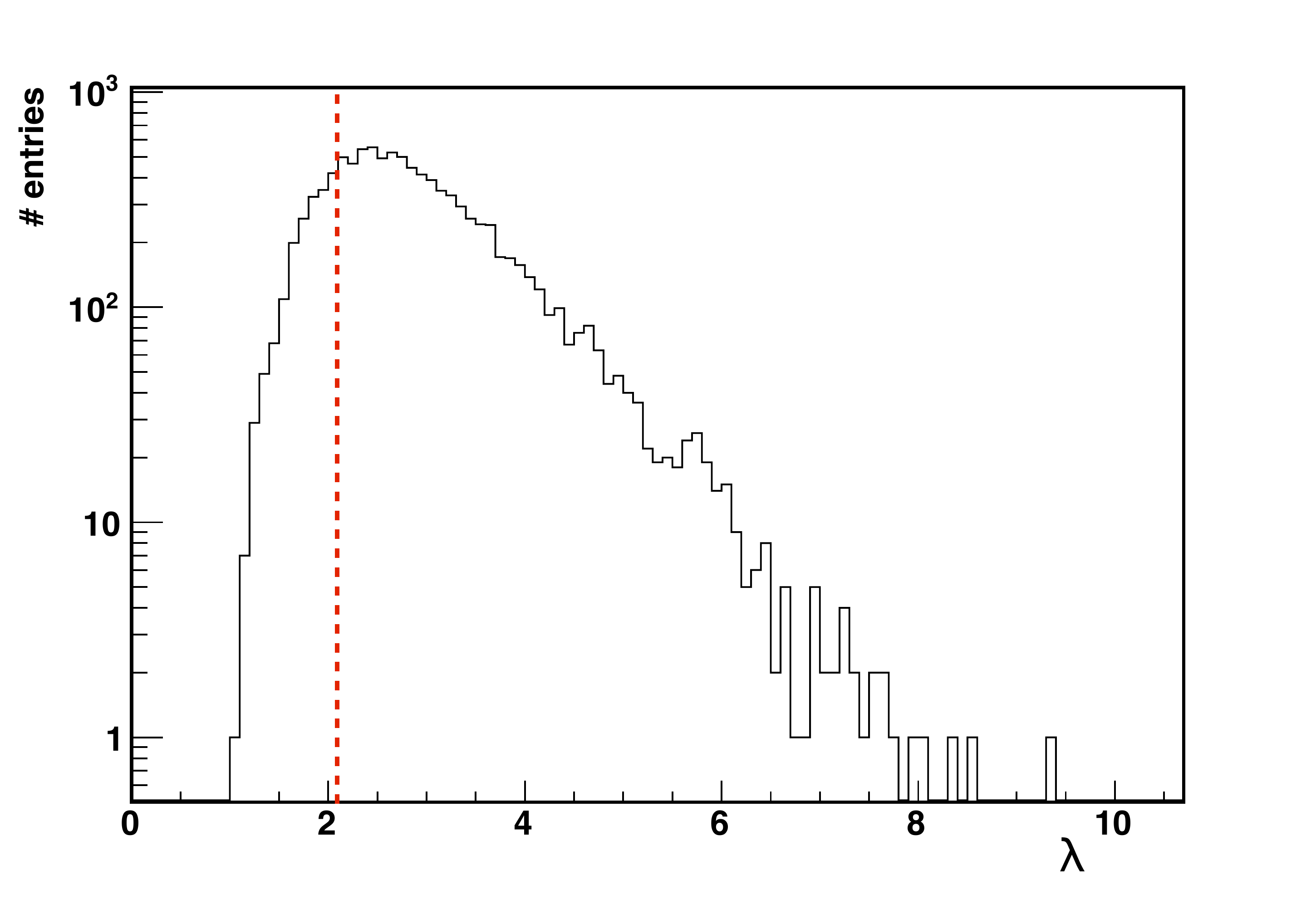}
 \caption{\label{fig:hotspots} Distribution of the largest value of $\lambda$ observed in each scrambled data set. The red dotted line indicated the value for $\lambda$ that corresponds to the hottest spot in the data.}
 \end{figure}
 
Figure \ref{fig:pointsources} displays a map of the sky with declination between $-85^{\circ}$ and $-60^{\circ}$ showing the events in this region and contours of the test statistic $\lambda$. The maximum value is $\lambda = 2.1$ at $\delta=-65.4^{\circ}$ and RA$=28.7^{\circ}$, corresponding to a fit of $n_s=3.5$ signal events. The overall significance of this value for $\lambda$ is found by producing $10,000$ scrambled data sets by randomizing the RA of each event. Figure \ref{fig:hotspots} shows the distribution of $\lambda$ associated to the hottest spot in each scrambled data set. The median test statistic value for the hottest spot in the scrambled data sets is $\lambda=2.7$, so the actual data set is consistent with a flat background. 

\begin{figure*}
 \includegraphics[width=\linewidth]{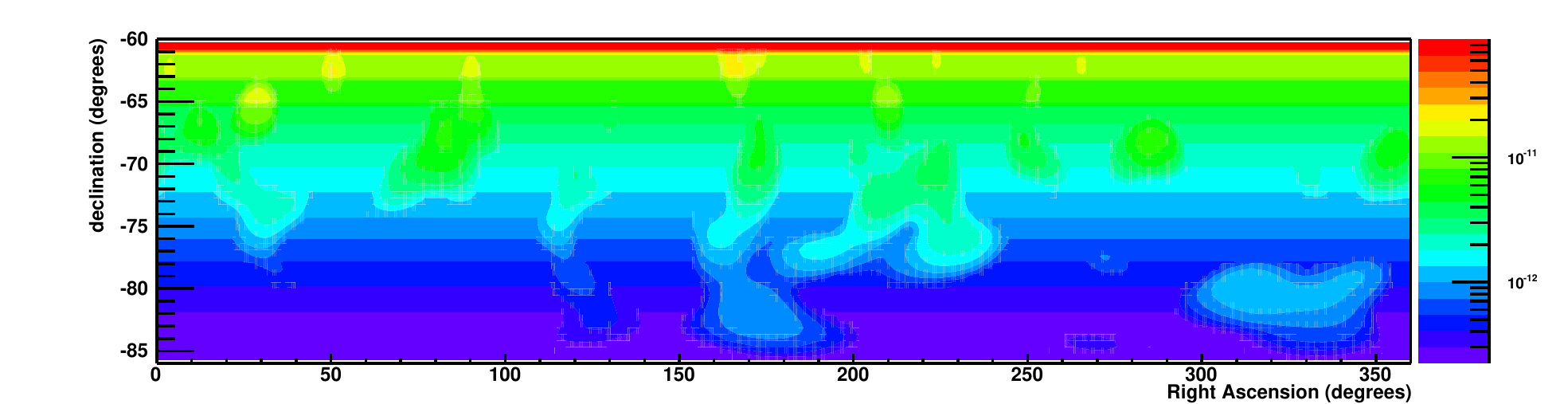}
 \caption{\label{fig:pointsourceslimits} Sky map of 90\% C.L. upper limits on point source flux $(E/\mathrm{TeV})^2 \mathrm{d}\Phi/\mathrm{d}E$ in cm$^{-2}$s$^{-1}$TeV$^{-1}$ for $E^{-2}$ source spectra in the energy range $E=1.2 - 6.0$ PeV. The limit is typically more constraining at low declinations where the effective area is largest. }
 \end{figure*}

Upper limits on the gamma-ray flux can be derived for each point in the sky by assuming that all events are gamma-rays. Since many events are in fact muon-poor or misreconstructed showers, this leads to a conservative upper limit. 
Because the acceptance of IC40 decreases as a function of declination (see Fig.~\ref{fig:acceptance}), the limit is more constraining at lower declination.
Figure  \ref{fig:pointsourceslimits} is a sky map of the 90\% C.L. upper limit in the energy range $E=1.2 - 6.0$ PeV for $E^{-2}$ source spectra. Point source fluxes are excluded at a level of $(E/\mathrm{TeV})^2 \mathrm{d}\Phi/\mathrm{d}E \sim 10^{-12} - 10^{-11}$ cm$^{-2}$s$^{-1}$TeV$^{-1}$ depending on source declination. Corrections for signal efficiency and detector noise are taken into account. 
Systematic uncertainties lead to a 18\% variation of the upper limit.

\section{Systematic Errors}
\label{sec:systerror}

Since this analysis derives the background from data, the systematic uncertainties due to the background estimation are small. The previously discussed cosmic-ray anisotropy measurement (see Sec.~\ref{sec:principle} and \cite{ic:anisotropy2}) is too small to have an impact on this analysis. Since there are no systematic gaps in detector uptime with respect to sidereal time-of-day in our sample, the coverage of RA is homogeneous.
   
Therefore, we focus on the systematic uncertainties in the signal efficiency, due to uncertainties in the surface detector sensitivity, and in the muon production rate for photon showers.    

The uncertainty in the surface detector sensitivity is studied in Ref. \cite{IceCube:2012wn}; Table 2 there gives the uncertainties for hadronic showers as a function of shower energy and zenith angle. Although there are differences between hadronic and electromagnetic showers, most factors that contribute to this figure apply to both types of showers. Strongly contributing factors include atmospheric fluctuations, calibration stability and uncertainties in response of detector electronics (PMT saturation and droop). The contribution from the uncertainty in modeling the hadronic interaction is clearly different for electromagnetic showers, and is discussed below. For $E<10$~PeV, and zenith angle less than 30$^{\circ}$, there is a 6.0\% systematic uncertainty in energy, and a 3.5\% systematic uncertainty in flux. For an $E^{-2.7}$ spectrum, a 6.0\% uncertainty in energy translates into a 17.0\% uncertainty in flux, or, adding in quadrature, 17.4\% flux uncertainty.

The uncertainty in the muon production from hadronic showers emerges from theoretical uncertainties.  It depends on the hadronic photoproduction and electroproduction cross-sections for energies between 10 TeV and 6 PeV.  Figure 1 of \cite{Couderc:2009tq} compares two cross-sections from two different photoproduction models, and finds (for water with a similar atomic number and mass number as air), a difference that rises from about 20\% at 10 TeV to 60\% at 1 PeV.  
The bulk of the particles in the shower are at lower energies, so we adopt a 20\% uncertainty on the muon production rate via photoproduction.
In addition, there is also a contribution of muon pair creation. To reach the in-ice DOMs, muons need at least 500 GeV. At 1~TeV, the fractional contribution of muon pair creation is $\sim 10$\% \cite{Stanev85}. Since muon pair production is not included in SYBILL 2.1, we arrive at 30\% uncertainty in total muon production rate.   
This uncertainty is applied to the 16\% of photon showers that are lost because they contain muons for a final 4.8\% uncertainty in sensitivity due to the unknown muon production cross-section.  

We add the uncertainties due to detector response and muon production in quadrature, and arrive at an overall 18 \% uncertainty in sensitivity. 

\section{IceCube 5-year sensitivity}
\label{sec:fullsensitivity}
The sensitivity of the full IceCube detector to a gamma-ray flux from the Galactic Plane benefits from multiple improvements that can be made with respect to the analysis presented above. 
In this section we use preliminary data from the IC79 configuration (79 strings,
73 surface stations, 2010/2011) to estimate the sensitivity that the
full IceCube detector can reach in 5 years. 
Since the full detector (IC86: 86 strings and 81 surface stations) is slightly larger than the IC79 configuration, the predicted sensitivity will be slightly underestimated. Also, the new cuts proposed below are not yet optimized, as this would require the actual IC86 data set.    
 
\subsection{Air shower reconstruction}

This analysis is very sensitive to the quality of the core reconstruction. If the shower core is not reconstructed accurately, a muon bundle that passes outside the in-ice array might be incorrectly assumed to be aimed at the detector. Because of the absence of a signal in the in-ice DOMs, the event is then misinterpreted as a gamma-ray candidate.
A more accurate core reconstruction algorithm has been developed for IceTop and will improve the CR rejection in post-IC40 analyses. In addition, the angular resolution of the larger array is improved, increasing the sensitivity to point sources.

\subsection{Isolated hits}
The SLC mode (which is available since IC59, see Sec.~\ref{sec:detector}) increases the sensitivity to CR showers with low muon content. A muon with just enough energy to reach IceCube, might not emit enough Cherenkov light to trigger multiple neighboring DOMs. By tightening the L4 cut so that no SLC hits are allowed to be present in the data, the efficiency with which CR showers can be rejected increases. At the same time, actual gamma-ray showers may be rejected in case of a noise hit in a single DOM. To keep this chance low, SLC hits only count as veto hits if they can be associated to the shower muon bundle both spatially and temporally.
 \begin{figure*}
 \includegraphics[width=0.48\linewidth]{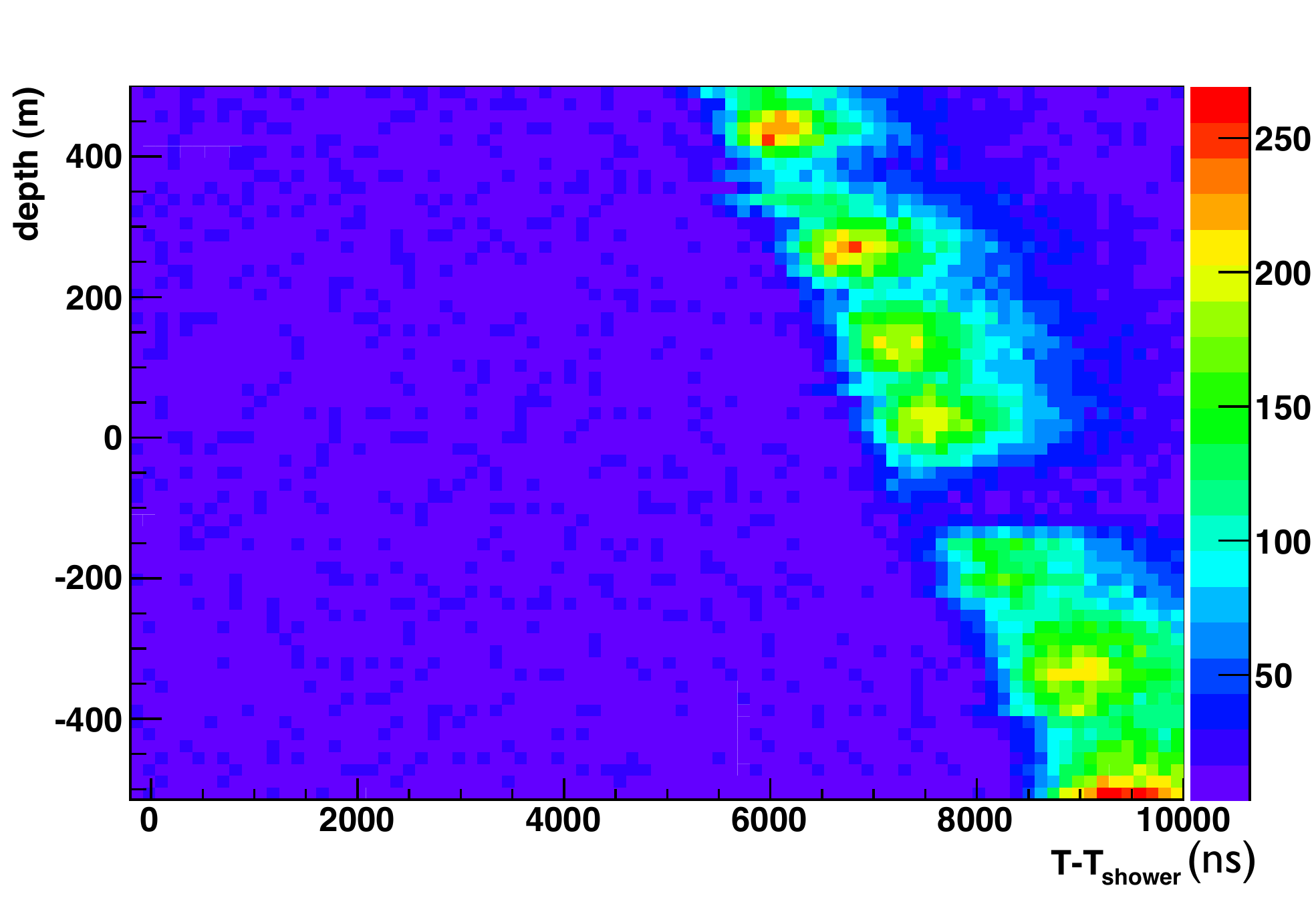}
 \includegraphics[width=0.48\linewidth]{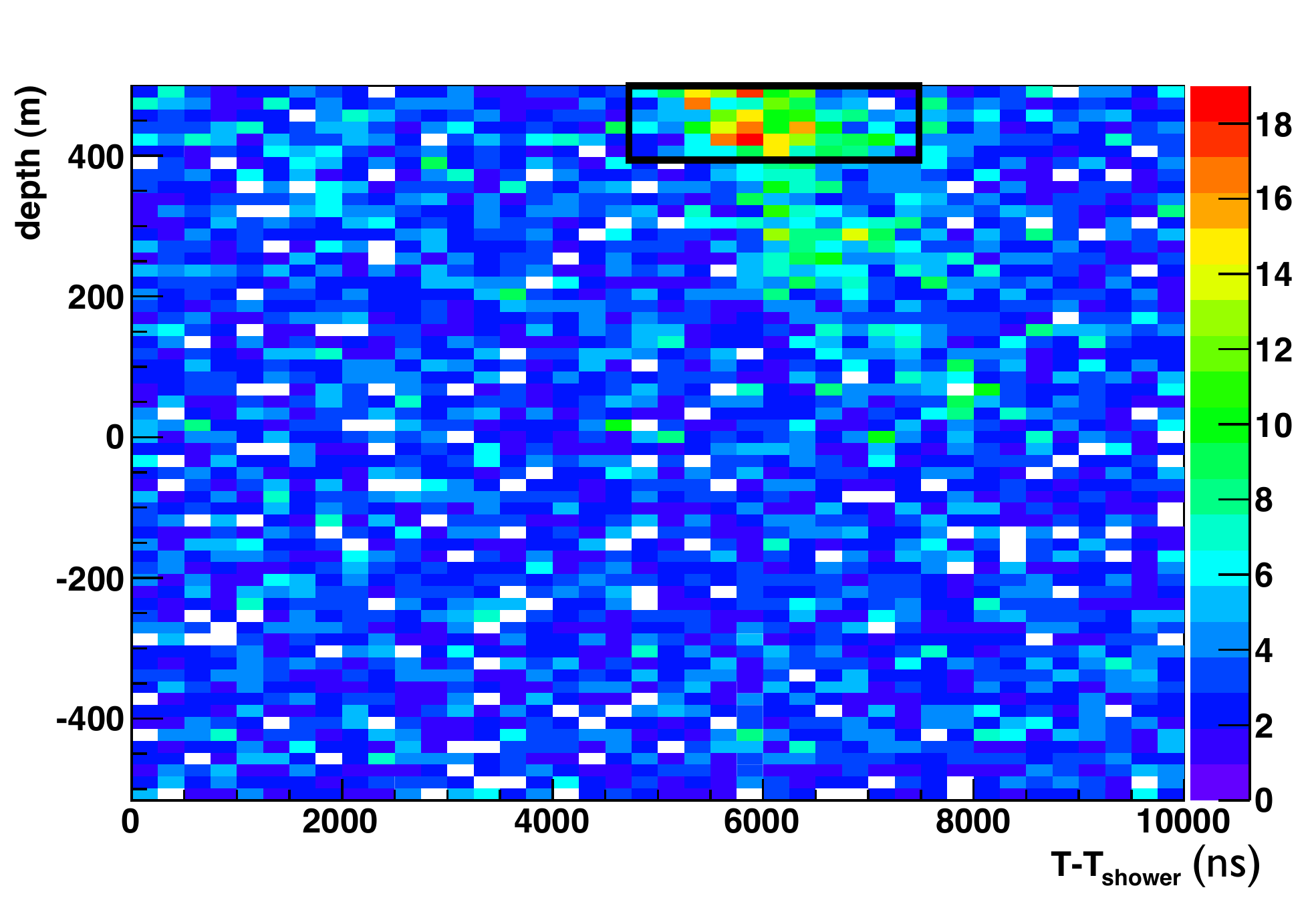}
 \caption{\label{fig:slchits} Observed time of isolated hits (SLC) relative to the arrival time of the air shower front measured by IceTop. The left plot shows the distribution of SLC hits for all events; the right plot is the same but restricted to the subset of events which have only SLC hits and no HLC hits. There is an excess of SLC hits in the region were a muon signal associated to the shower is expected. This allows an additional cut to separate gamma-ray showers from hadronic showers. The black box indicates the region in which an SLC hit counts as a veto (see text for details). The variation in the number of hits as a function of depth in the left plot is due to variations in the optical properties of the ice.}
 \end{figure*}
 
Figure \ref{fig:slchits} shows the distribution of isolated hits in the complete detector as a function of time relative to the arrival time of the air shower as measured by IceTop. The plots show data at L3 level, applying the same cut values as in the IC40 analysis.
The left plot shows the distribution of SLC hits for all events, while the right plot shows the same distribution but restricted to the subset of events which contain {\it only} SLC hits, i.e. events with no HLC hits.
Hits associated with the muon bundle are seen throughout the detector, although the number of hits varies with depth because of 
variations in the optical properties of the ice due to naturally varying levels of contaminants such as dust, which attenuate Cherenkov photons 
The large number of isolated hits in the two bottom rows is an edge effect: the DOMs have fewer neighbors, so the chance for a hit to be isolated increases. In principle, the same effect could occur at the top two rows. However, the muon bundle deposits more energy in this region and the probability for any hit to have neighbor hits is larger here.   

The muon-poor showers that produce no HLC hits (right-hand plot) can still cause some isolated hits in the top of the detector. These events can be removed with an additional cut on SLC hits. Because isolated hits can also be produced by noise, only a small area is selected in which SLC hits are used as a veto. A simple additional L4 cut is that all events are removed that have an SLC hit meeting the following three criteria:
\begin{itemize}
\item it is within 200 m from the reconstructed shower axis,
\item it is within a time window of 4.8--7.5 $\mu$s after the shower arrival time, and
\item it is in one of the six top layers of DOMs (spanning a vertical extend of 85~m).
\end{itemize}
Note that the lower bound of the time window (4.8  $\mu$s) corresponds to the time it takes for a muon traveling vertically to reach the top layer of in-ice DOMs starting from the surface. Muons from an inclined shower will arrive even later.
The number of background events that are discarded in the L4 cut is increased by $\sim$30\%, while the SLC noise rate in the data implies a decrease in signal efficiency of $\sim$5\%.

With the completed detector it will be possible to optimize the SLC cuts further by making the time window dependent on zenith angle and DOM depth. The effect of this optimization was not yet studied here.

\subsection{Re-optimization of cuts}

 \begin{figure*}
 \includegraphics[width=0.3\linewidth]{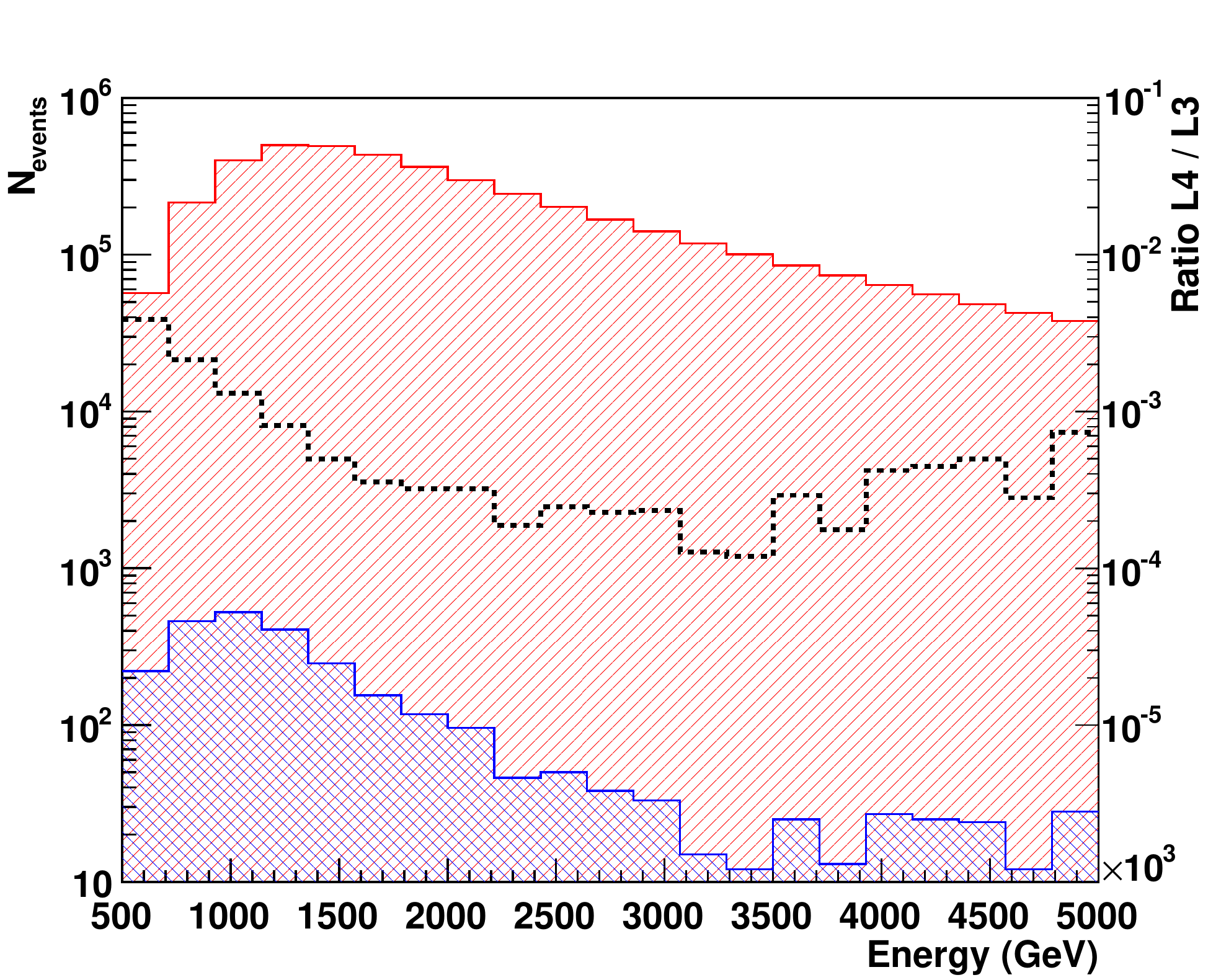}
 \includegraphics[width=0.3\linewidth]{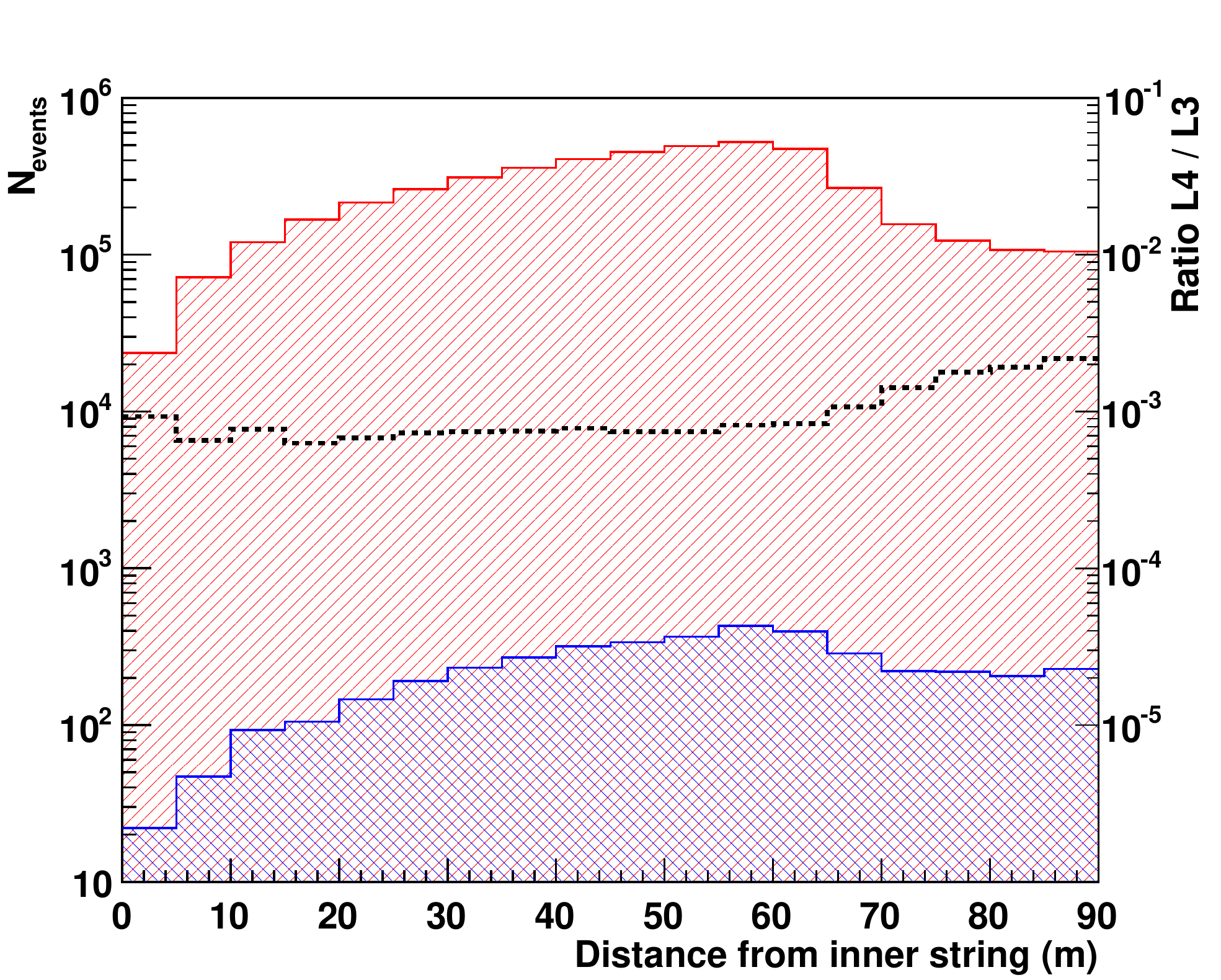}
 \includegraphics[width=0.3\linewidth]{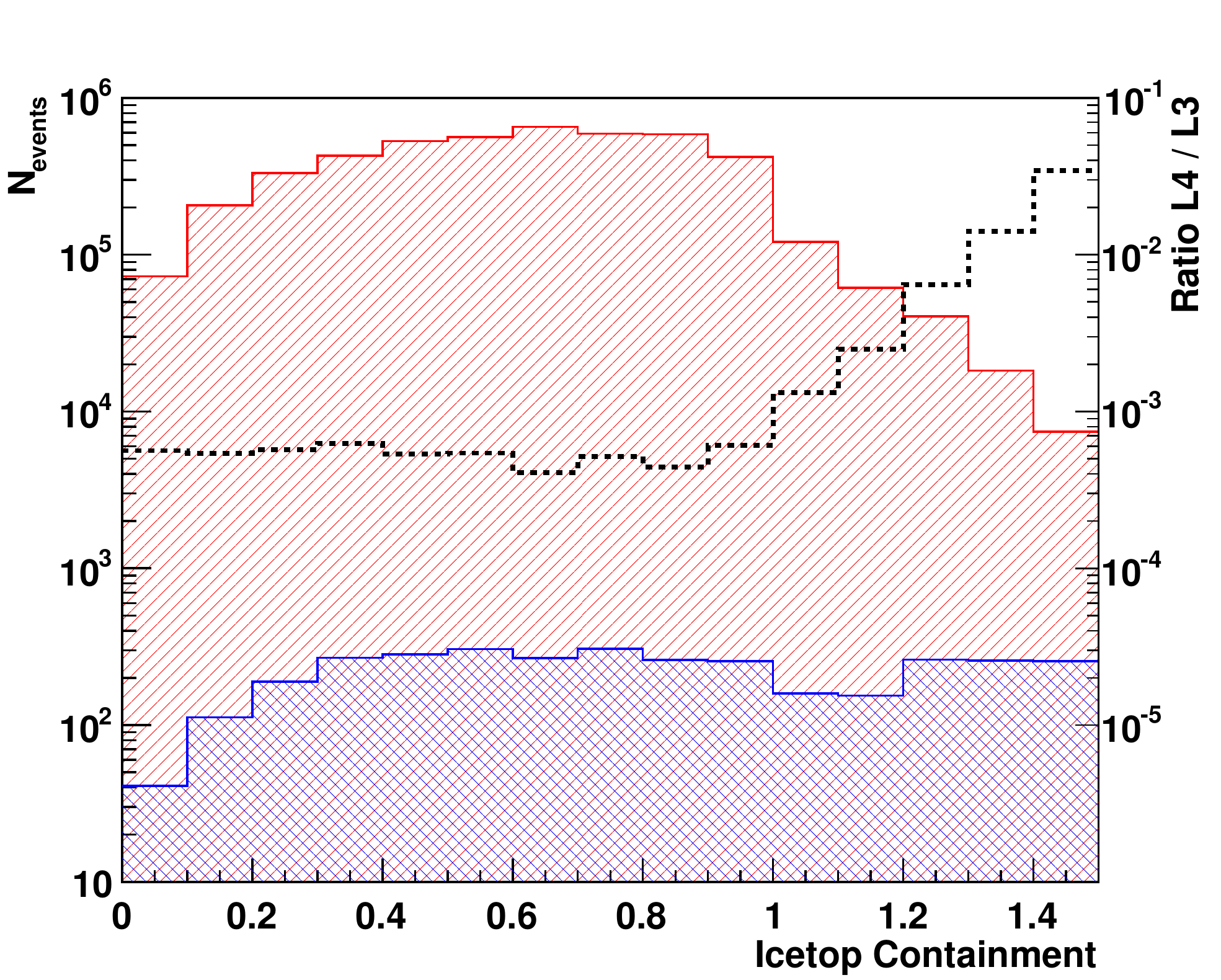}
 \caption{\label{fig:ratios} The number of L3 (red) and L4 (blue) events in the data as a function of the three main cut parameters. The ratio of the number of L4 to L3 events is given by the black dotted line, for which the corresponding axis is drawn on the right-hand side of the plot.}
 \end{figure*}

For the IC40 analysis the cut parameters were optimized to increase the detection probability of a possible correlation of gamma rays with the Galactic Plane. 
To increase the sensitivity of future searches with the completed IC86 configuration, the cut values were re-evaluated to increase the number of candidate events without losing background rejection power. This was achieved by evaluating the ratio between the number of events after L3 and L4 cuts. 

While the L3 event sample is completely dominated by CRs, the L4 sample is a combination of possible gamma-ray showers, muon-poor CR showers and misreconstructed CR showers. The fraction of gamma-rays and muon-poor CRs in the detected events is independent from the cuts on geometry parameters $d_{\mathrm{str}}$ and $C_{\mathrm{IT}}$. The number of misreconstructed CR showers, on the other hand, will increase if the geometry cut values are chosen too loosely. Therefore, the ratio between the number of L4 and L3 events as a function of the cut parameter should be flat up to some maximum value after which it starts to increase. This maximum value is the preferred cut value since it maximizes the number of candidate events without lowering the background rejection power. It also maximizes the FOV, as looser geometry cuts imply a larger maximum zenith angle.

Figure \ref{fig:ratios} shows the number of L3 (red) and L4 (blue) events together with their ratio (black dotted line; right-hand axis) as a function of the three main cut parameters (with the other cut parameters kept constant at their final value).  

The rejection efficiency for $d_{\mathrm{str}}$ is fairly stable up to 60~m. The number of events rapidly decreases above this value, while the rejection becomes worse. In this case, the alternative method of optimization yields the same result as the method used in the IC40 analysis. For the containment size $C_{\mathrm{IT}}$ the ratio remains stable up to the edge of the array ($C_{\mathrm{IT}}=1$) after which it starts to rise. It appears the cut can be relaxed with respect to the IC40 analysis. In the following we will use $d_{\mathrm{str}}<60$~m and $C_{\mathrm{IT}}<1.0$.

The efficiency of the energy cut increases, as expected, with increasing energy, leveling off around $\sim$ 2.0 PeV. Since the total number of events falls off rapidly for increasing energy, the most sensitive region will be $\sim 2-3$ PeV. However, since the spectra of possible sources in this energy regime are unknown, it is not clear what energy cut would produce the optimal sensitivity. Instead, the sensitivity is calculated for ten energy bins in the range 1--10 PeV (see Fig.~\ref{fig:sensgalp}). 

\subsection{Increased acceptance}
\begin{figure}
\includegraphics[width=\linewidth]{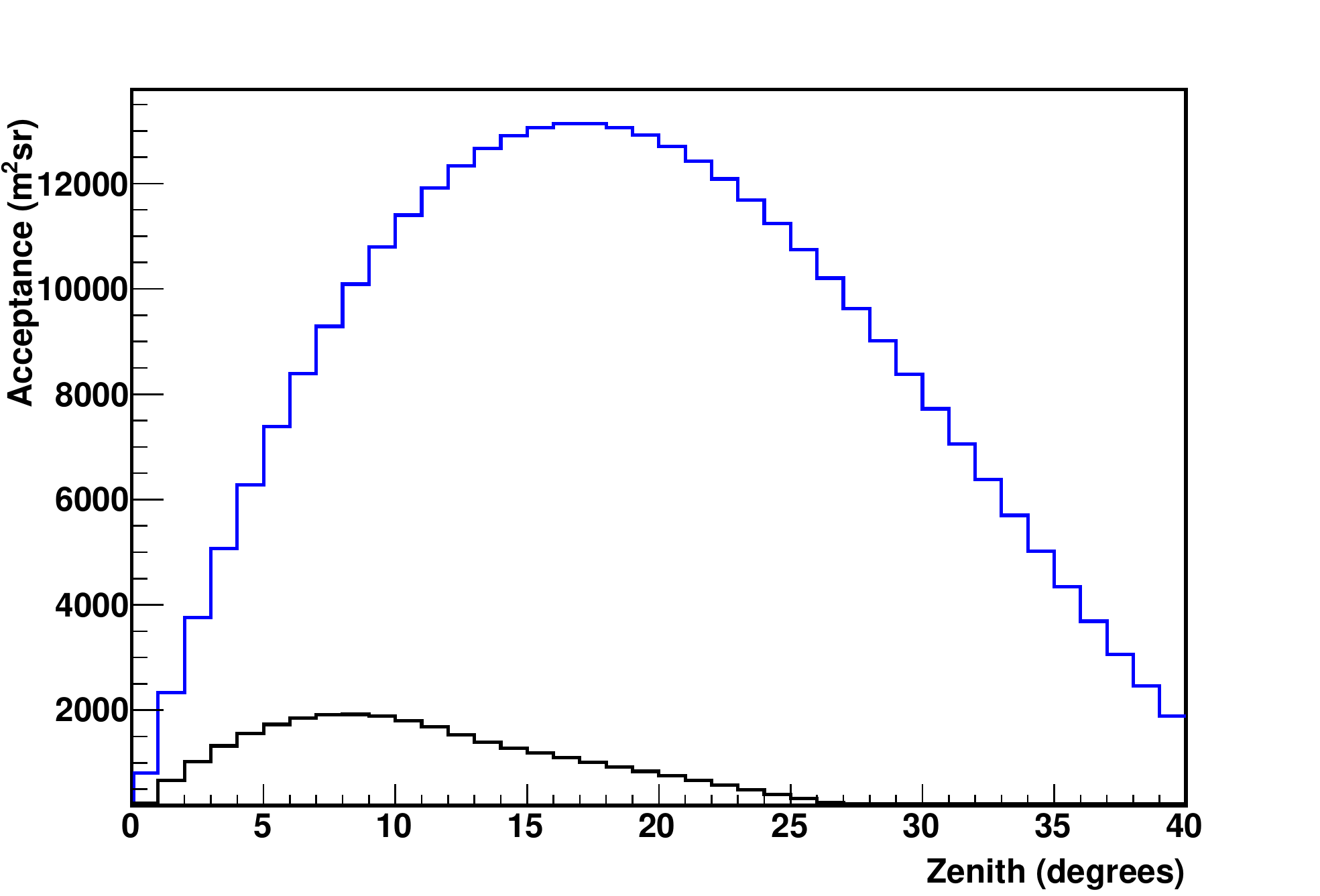}
\caption{\label{fig:acceptance} Acceptance (effective area integrated over solid angle) for showers with an axis through both IceTop and IceCube for IC40, $C_{\mathrm{IT}}$=0.8 (black), and IC86, $C_{\mathrm{IT}}$=1.0 (blue).}
\end{figure}

With a larger array the acceptance, defined here as the effective area integrated over the solid angle of each 1$^{\circ}$ bin in zenith angle, increases considerably. Because of the condition that the shower axis has to be inside the instrumented area of both IceCube and IceTop, the increase is especially dramatic at larger zenith angles. Fig.~\ref{fig:acceptance} shows the acceptance for IC40 with $C_{\mathrm{IT}} < 0.8$ and the complete IC86 array with $C_{\mathrm{IT}} < 1.0$. Not only does the acceptance increase at large zenith angles, the range of possible zenith angles is also extended (to  $\approx 45^{\circ}$). This extends the FOV to cover a larger part of the Galactic Plane and probe an area closer to the Galactic Center. The Galactic Center itself is still outside the FOV at $\delta \approx -29^{\circ}$, corresponding to a zenith angle of $61^{\circ}$.

\subsection{Sensitivity}
\begin{figure}
\includegraphics[width=\linewidth]{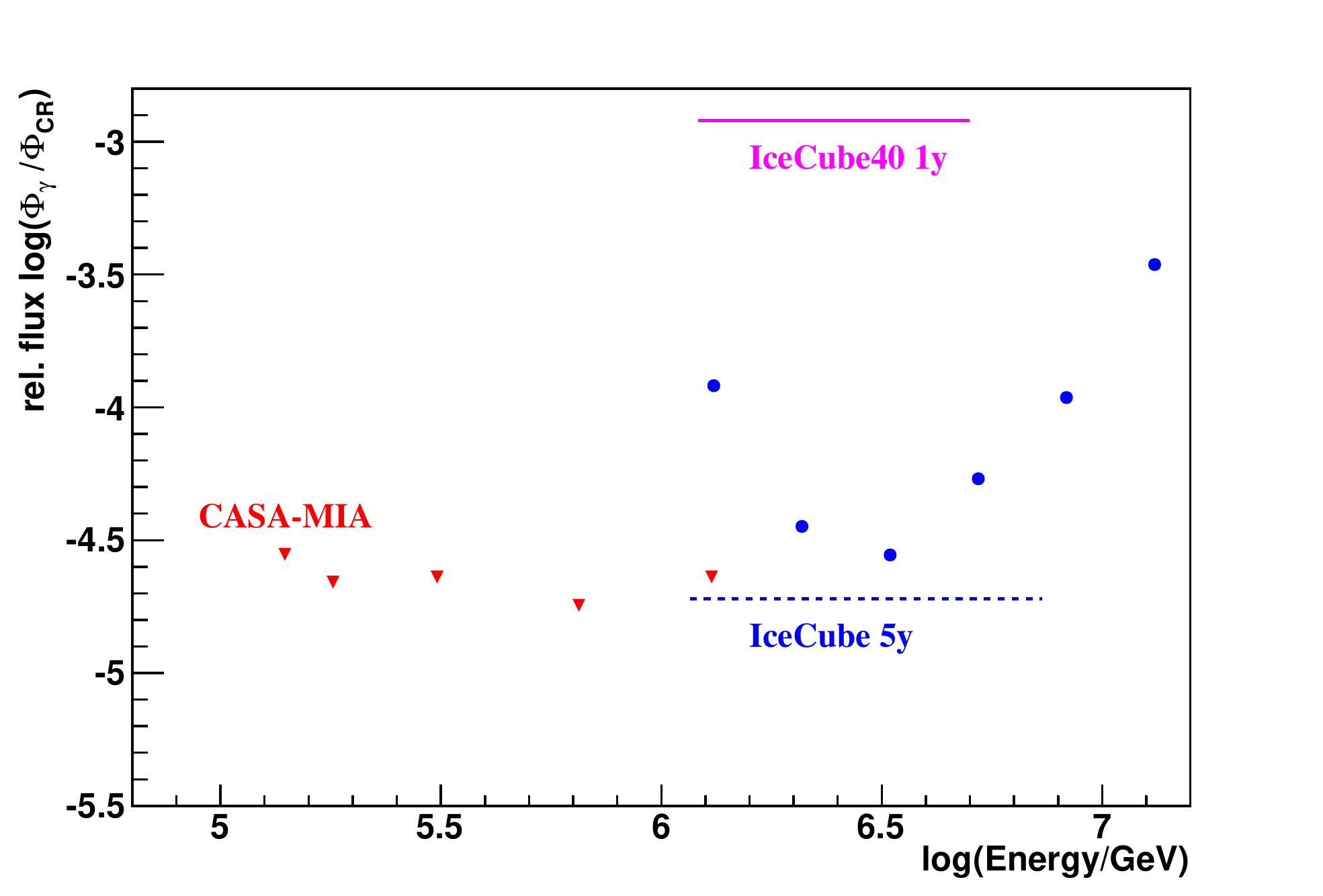}
\caption{\label{fig:sensgalp} Existing limits (red triangles for CASA-MIA and purple line for present IC40 analysis) and IceCube sensitivity to a diffuse gamma-ray flux from a region within 10 degrees from the Galactic Plane. The blue dashed line indicates the five year sensitivity of the completed detector, while the blue dots represent the sensitivity in smaller energy bins.}
\end{figure}

The sensitivity that can be reached with 5 years of data from the completed IceCube configuration can be estimated with preliminary data from IC79.
It is assumed that the fraction of gamma-rays that are missed due to noise hits is the same as in the IC40 analysis. The full detector obviously has more noise hits, but this can be compensated by refining the in-ice cut by only allowing vetoes from DOMs that can be associated to the shower muon bundle in space and time (cf.~the SLC cut described above). The sensitivity is calculated by producing scrambled data sets with randomized RA. Figure \ref{fig:sensgalp} shows the 90\% C.L. sensitivity to a diffuse flux from within 10 degrees of the Galactic Plane that can be achieved with 5 years of full detector data. The blue dashed line indicates the integrated limit between 1 and 10 PeV, while the blue dots indicate the sensitivity in six smaller energy bins. The upper limits found by CASA-MIA and IC40 (present work) are also included in the plot. The KASCADE \cite{KASCADE} results are not included since they set a limit on the all-sky gamma-ray flux. 

\begin{figure}
\includegraphics[width=\linewidth]{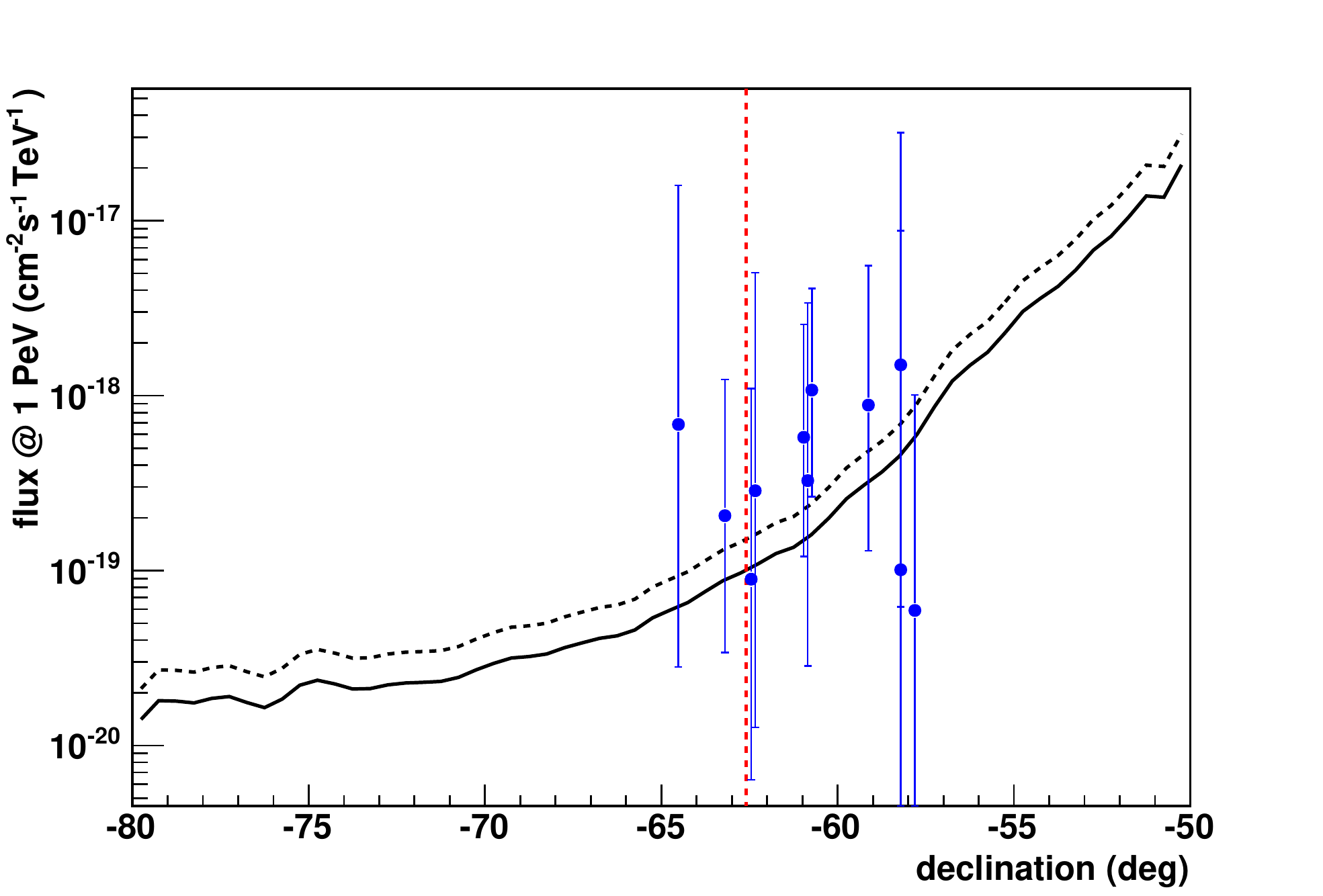}
\caption{\label{fig:sensps} IceCube 5 year sensitivity to point sources as a function of declination. The solid (dashed) black line indicates the sensitivity to an $E^{-2} (E^{-2.5}) $ flux. The dashed red line indicates the lowest declination reached by the Galactic Plane. The blue points indicate the flux at 1 PeV with extrapolated uncertainties of the sources listed in Table \ref{table:sources} in the absence of a cut-off.}
\end{figure}

Figure~\ref{fig:sensps} shows the sensitivity to point sources that is possible with 5 years of IceCube data.
The sensitivity is a strong function of declination because the acceptance decreases at larger zenith angles.
Point sources are expected to lie close to the Galactic Plane which reaches its lowest declination at $-63^{\circ}$. Within the IceCube field of view there are several PWNe and other gamma-ray sources detected by H.E.S.S.  \cite{HESS}, listed in Table  \ref{table:sources}. 
For these sources no significant cut-off was observed up to the maximum energy of 10 TeV, where statistics gets low.
The blue dots indicate the flux that these sources would have at 1 PeV if their spectrum remains unchanged up to that energy. 
No correction for gamma-ray attenuation  between the source and observer has been applied in this calculation.
The extrapolation over two order of magnitude causes large uncertainties in the gamma-ray flux due to propagation of the errors on the spectral indices.

\begin{table*}
\caption{\label{table:sources} List of H.E.S.S. sources in IceCube gamma-ray FOV. For those values that have two error margins, the first indicates the statistical error, while the second indicates the systematic error. }
\begin{ruledtabular}
\begin{tabular}{llllll}
Source & RA & decl. & Flux at 1 TeV (cm$^{-2}$s$^{-1}$TeV)$^{-1})$& $\Gamma$ & Classification\\
\hline
HESS J1356-645 & $13^{h}56^{m}00^{s}$ & $-64^{\circ}30^{\prime}00^{\prime\prime}$ &$(2.7\pm0.9\pm0.4) \times 10^{-12}$&$ 2.2\pm 0.2 \pm 0.2$ & PWN \cite{src:J1356-645}\\
HESS J1303-631 & $13^{h}03^{m}00^{s}$ & $-63^{\circ}11^{\prime}55^{\prime\prime}$ &$(4.3\pm0.3) \times 10^{-12}$&$ 2.44\pm 0.05 \pm 0.2$ & PWN \cite{src:J1303-631} \\
RCW 86 &                 $14^{h}42^{m}43^{s}$ & $-62^{\circ}28^{\prime}48^{\prime\prime}$ &$(3.72\pm0.5\pm0.8) \times 10^{-12}$&$ 2.54\pm 0.12 \pm 0.2$ & shell-type SNR \cite{src:RCW86} \\
HESS J1507-622 & $15^{h}06^{m}53^{s}$ & $-62^{\circ}21^{\prime}00^{\prime\prime}$ &$(1.5\pm0.4\pm0.3) \times 10^{-12}$&$ 2.24\pm 0.16 \pm 0.2$ & no ID \cite{src:J1507-622} \\
Kookaburra (Rabbit) & $14^{h}18^{m}04^{s}$ & $-60^{\circ}58^{\prime}31^{\prime\prime}$ &$(2.64\pm0.2\pm0.53) \times 10^{-12}$&$ 2.22\pm 0.08\pm 0.1$ & PWN \cite{src:kookaburra}\\
HESS J1427-608 & $14^{h}27^{m}52^{s}$ & $-60^{\circ}51^{\prime}00^{\prime\prime}$ &$(1.3\pm0.4) \times 10^{-12}$&$ 2.2\pm 0.1 \pm 0.2$ & no ID \cite{src:J1427-608}\\
Kookaburra (PWN) & $14^{h}20^{m}09^{s}$ & $-60^{\circ}45^{\prime}36^{\prime\prime}$ &$(3.48\pm0.2\pm0.7) \times 10^{-12}$&$ 2.17\pm 0.06\pm 0.1$ & PWN \cite{src:kookaburra} \\
MSH 15-52 &            $15^{h}14^{m}07^{s}$ & $-59^{\circ}09^{\prime}27^{\prime\prime}$ &$(5.7\pm0.2\pm1.4) \times 10^{-12}$&$ 2.27\pm 0.03 \pm 0.2$ & PWN \cite{src:MSH15-52}\\
HESS J1503-582 & $15^{h}03^{m}38^{s}$ & $-58^{\circ}13^{\prime}45^{\prime\prime}$ &$(1.6\pm0.6) \times 10^{-12}$&$ 2.4\pm 0.4\pm 0.2$ & dark (FWV?) \cite{src:J1503-582}\\
HESS J1026-582 & $10^{h}26^{m}38^{s}$ & $-58^{\circ}12^{\prime}00^{\prime\prime}$ &$(0.99\pm0.34) \times 10^{-12}$&$ 1.94\pm 0.2\pm 0.2$ & PWN \cite{src:westerlund2} \\
Westerlund 2 & $10^{h}23^{m}24^{s}$ & $-57^{\circ}47^{\prime}24^{\prime\prime}$ &$(3.25\pm0.5) \times 10^{-12}$&$ 2.58\pm 0.19 \pm 0.2$ & MSC \cite{src:westerlund2}\\

\end{tabular}
\end{ruledtabular}
\end{table*}

\section{Conclusions}
We have presented a new method of searching for high energy gamma-rays using the IceCube detector and its surface array IceTop. 
One year of data from IC40 was used to perform a search for point sources and a Galactic diffuse signal. No sources were found, resulting in a 90\% C.L. upper limit on the ratio of gamma rays to cosmic rays of $1.2\times 10^{-3}$ for the flux coming from 
the Galactic Plane region ( $-80^\circ \lesssim l \lesssim -30^\circ; -10^\circ \lesssim b \lesssim 5^\circ$)
in the energy range 1.2 -- 6.0 PeV. Point source fluxes with $E^{-2}$ spectra have been excluded at a level of $(E/\mathrm{TeV})^2 \mathrm{d}\Phi/\mathrm{d}E \sim 10^{-12} - 10^{-11}$ cm$^{-2}$s$^{-1}$TeV$^{-1}$ depending on source declination. The full detector was shown to be much more sensitive, because of its larger size, improved reconstruction techniques and the possibility to record isolated hits. 

This analysis offers interesting observation possibilities. IceCube can search for a diffuse Galactic gamma-ray flux with a sensitivity comparable to CASA-MIA, but at higher energies. This sensitivity is reached, however, by studying a much smaller part of the Galactic Plane than CASA-MIA. IceCube is therefore especially sensitive to localized sources, which might be Galactic accelerators or dense targets for extragalactic CRs. 

The H.E.S.S.\ and CANGAROO-III \cite{CANGAROO} telescopes have found several high energy gamma-ray sources in IceCube's FOV. Most of these sources are identified as or correlated with PWNe. Their energy spectrum has been measured up to a couple of tens of TeV. At this energy, statistics become low and for most sources no cut-off has been established.
If these spectra extend to PeV energies without a break, IceCube will be able to detect them. It is also possible that an additional spectral component in the PeV energy range is present if a nearby dense molecular cloud acts as a target for the PWN beam \cite{Bednarek}. 
IceCube will be able to study these systems and place constraints on their behavior at very high energies, or possibly detect PeV gamma-rays for the first time. 
   
\begin{acknowledgments}

We acknowledge the support from the following agencies:
U.S. National Science Foundation-Office of Polar Programs,
U.S. National Science Foundation-Physics Division,
University of Wisconsin Alumni Research Foundation,
the Grid Laboratory Of Wisconsin (GLOW) grid infrastructure at the University of Wisconsin - Madison, the Open Science Grid (OSG) grid infrastructure;
U.S. Department of Energy, and National Energy Research Scientific Computing Center,
the Louisiana Optical Network Initiative (LONI) grid computing resources;
National Science and Engineering Research Council of Canada;
Swedish Research Council,
Swedish Polar Research Secretariat,
Swedish National Infrastructure for Computing (SNIC),
and Knut and Alice Wallenberg Foundation, Sweden;
German Ministry for Education and Research (BMBF),
Deutsche Forschungsgemeinschaft (DFG),
Research Department of Plasmas with Complex Interactions (Bochum), Germany;
Fund for Scientific Research (FNRS-FWO),
FWO Odysseus programme,
Flanders Institute to encourage scientific and technological research in industry (IWT),
Belgian Federal Science Policy Office (Belspo);
University of Oxford, United Kingdom;
Marsden Fund, New Zealand;
Australian Research Council;
Japan Society for Promotion of Science (JSPS);
the Swiss National Science Foundation (SNSF), Switzerland.

\end{acknowledgments}


\end{document}